\pdfoutput=1
\documentclass[aps, onecolumn, showpacs, superscriptaddress, prx]{revtex4-2}

\usepackage[pdftex]{graphicx}
\usepackage[pdftex]{epsfig}
\usepackage{epstopdf}
\usepackage{color}
\usepackage{amssymb,amsmath}
\usepackage[hypertexnames=false,colorlinks,urlcolor=blue,citecolor=blue]{hyperref}
\usepackage{braket}
\usepackage{siunitx}
\usepackage[capitalise]{cleveref}
\usepackage{float}

\usepackage[dvipsnames]{xcolor}

\newcommand{\proj}{\mathrm{proj}}
\newcommand{\elec}{\mathrm{elec}}

\newcommand{\norm}[1]{\left\lVert#1\right\rVert}

\newcommand{\sel}{\textsc{sel}}
\newcommand{\prep}{\textsc{prep}}
\newcommand{\nn}{\nonumber \\}

\usepackage[english]{babel}

\usepackage{algorithm2e}

\newcommand{\MQ}{\affiliation{%
School of Mathematical and Physical Sciences,
Macquarie University, Sydney, NSW, AUS}}

\newcommand{\Google}{\affiliation{
Google Quantum AI, Venice, CA, USA}}

\newcommand{\Harvard}{\affiliation{Department of Chemistry and Chemical Biology, Harvard University, Cambridge, MA, USA}}

\newcommand{\SandiaNM}{\affiliation{Quantum Algorithms and Applications Collaboratory, Sandia National Laboratories, Albuquerque, NM, USA}}

\newcommand{\QSimulate}{\affiliation{%
Quantum Simulation Technologies Inc., Boston, MA, USA}}

\begin{document}

\title{Quantum computation of stopping power for inertial fusion target design}

\author{Nicholas C.~Rubin}
\email[Corresponding author: ]{nickrubin@google.com}
\Google

\author{Dominic W.~Berry}
\email[Corresponding author: ]{dominic.berry@mq.edu.au}
\MQ

\author{Alina Kononov}
\SandiaNM

\author{Fionn D.~Malone}
\Google

\author{Tanuj Khattar}
\Google

\author{Alec White}
\QSimulate

\author{Joonho Lee}
\Google
\Harvard

\author{Hartmut Neven}
\Google

\author{Ryan Babbush}
\email[Corresponding author: ]{babbush@google.com}
\Google

\author{Andrew D. Baczewski}
\email[Corresponding author: ]{adbacze@sandia.gov}
\SandiaNM

\begin{abstract}
Stopping power is the rate at which a material absorbs the kinetic energy of a charged particle passing through it -- one of many properties needed over a wide range of thermodynamic conditions in modeling inertial fusion implosions. First-principles stopping calculations are classically challenging because they involve the dynamics of large electronic systems far from equilibrium, with accuracies that are particularly difficult to constrain and assess in the warm-dense conditions preceding ignition. Here, we describe a protocol for using a fault-tolerant quantum computer to calculate stopping power from a first-quantized representation of the electrons and projectile. Our approach builds upon the electronic structure block encodings of Su \emph{et al.}~\cite{Su2021}, adapting and optimizing those algorithms to estimate observables of interest from the non-Born-Oppenheimer dynamics of multiple particle species at finite temperature. Ultimately, we report logical qubit requirements and leading-order Toffoli costs for computing the stopping power of various projectile/target combinations relevant to interpreting and designing inertial fusion experiments. We estimate that scientifically interesting and classically intractable stopping power calculations can be quantum simulated with
roughly the same number of logical qubits and about one hundred times more Toffoli gates than is required for state-of-the-art quantum simulations of industrially relevant molecules such as FeMoCo or P450.
\end{abstract}

\maketitle

\section{Introduction}
\label{sec:introduction}

As investment in quantum computing grows, so too does the need to assess potential computational advantages for specific scientific challenge problems. 
As such, going beyond asymptotic analysis to understand constant factor resource requirements clarifies the degree of such advantages, whether these advantages would be useful for problems of commercial or scientific value, the broader quantum/classical simulation frontier, and opportunities for further optimizations. Over the last decade, the problem of sampling from the eigenspectrum of the electronic structure Hamiltonian in the Born-Oppenheimer approximation has been a proving ground for constant factor analyses that quantify the magnitude of quantum speedups~\cite{Lee2020, vonBurg2020, rubin2023fault, Berry2019QF}. But ground states can benefit from properties like area-law entanglement that might make many instances relevant to chemical and materials science efficient to accurately simulate in some contexts -- even with classical algorithms~\cite{lee2023evaluating}.  This leads us to search for scientific challenge problems beyond ground states, for which classical algorithms might have less structure to exploit and quantum algorithms are naturally poised to excel.  Simulating quantum dynamics is arguably the most natural application for quantum computers. Just as in the case of sampling from the eigenbasis of the electronic structure Hamiltonian, here we quantify the constant factor resource estimates of dynamics calculations in order to frame the performance of current quantum algorithms with respect to classical strategies. We focus on the problem of computing the stopping power of materials in the warm dense matter (WDM) regime, for which both experimental measurements and benchmark-quality theoretical calculations are expensive and sparse. 
Even for mean-field levels of accuracy, high-performance-computing campaigns requiring at least hundreds of millions of CPU hours are invested annually in first-principles stopping power calculations in WDM~\footnote{Two of the coauthors (AK and ADB) have maintained allocations on capability-class supercomputers to do stopping power calculations for nearly a decade.
In their last year of allocations, this has involved 14 days of CPU time on a machine with approximately 300,000 processors.}.

Stopping power is the average force exerted by a medium (target) on an incident charged particle (projectile)~\cite{bragg1905alpha,bohr1913ii}.
This force depends on the material composition and conditions of the target and the charge and velocity of the projectile.
Stopping powers are crucial in contexts including radiation damage in space environments~\cite{fan1996shielding}, materials degradation in nuclear reactors~\cite{sand2019heavy}, certain cancer therapies~\cite{schardt2010heavy,newhauser2015physics,shepard2023electronic}, electron-~\cite{joy1996low} and ion-beam~\cite{kononov2021anomalous,kononov2022first} microscopies, and the fabrication and characterization of qubits based on color centers~\cite{wan2020large} or nuclear spins~\cite{asaad2020coherent,jakob2022deterministic}.
Understanding the impact of stopping power on these applications is facilitated by the relative ease of conducting experiments, and decades of effort have produced tables of experimentally measured stopping powers for targets in ambient conditions~\cite{berger1999estar,ziegler2010srim,montanari2017iaea}.
Such measurements require co-locating uniform samples of the target with a well-characterized, narrow bandwidth source of high-energy projectiles and a spectrometer with sufficient resolution to discern small relative energy losses.
This experimental setup is comparatively straightforward for a stable target, but it becomes incredibly challenging for a target at extreme pressure or temperature.

Warm dense matter (WDM)~\cite{graziani2014frontiers,dornheim2018uniform} is one such extreme regime, typified by strong Coulomb coupling and the simultaneous influence of thermal and degeneracy effects.
It arises in contexts ranging from astrophysical objects~\cite{booth2015laboratory,kritcher2022design,lutgert2022platform} and planetary interiors~\cite{saumon2004shock,nettelmann2008ab,benuzzi2014progress} to inertial confinement fusion (ICF) targets on the way to ignition~\cite{hu2015impact,yager2022overview,abu2022lawson,hurricane2023physics}.
Creating WDM conditions requires access to specialized experimental facilities~\cite{soures1996direct,nagler2015matter,falk2018experimental,sinars2020review,macdonald2023colliding} that produce short-lived and non-uniform samples with low repetition rates relative to experiments at ambient conditions.
These challenges are compounded by the difficulty of simultaneously characterizing the target's thermodynamic conditions and the projectile's energy loss, as well as systematic errors attendant to measuring aggregate energy losses in lieu of energy loss rates.
Despite outstanding recent advances in measurements of stopping power in WDM~\cite{frenje2015measurements,zylstra2015measurement,malko2022proton}, theoretical calculations will likely remain disproportionately impactful due to the great cost of obtaining comprehensive data sets purely through experiment.
Each campaign typically probes stopping powers over a narrow range of velocities for a narrow range of thermodynamic conditions that might themselves be difficult to constrain or subject to large systematic uncertainties and inhomogeneities.

The importance of stopping power in WDM, in particular, is highlighted by its significance to ICF~\cite{zylstra2022burning}.
The transport of the high-energy alpha particles that are created in fusion reactions forms an important contribution to the self-heating processes that govern ignition~\cite{singleton2008charged,temporal2017effects,zylstra2019alpha}.
WDM conditions are necessarily traversed in ICF implosions \--- in fact, depending on the target and driver, large fractions of the target can spend most of their time in this regime.
This intermediate state also plays a central role in the fuel/ablator mixing that leads to hydrodynamic instabilities limiting performance~\cite{smalyuk2019review,zhou2019turbulent,gomez2020performance}.
Fast-ignition fusion concepts also rely on charged particle stopping within their separate ion or electron beam heating mechanism~\cite{roth2001fast,atzeni2008stopping,solodov2008stopping}.
Thus accurate stopping powers are one among many important elements of the microphysics modeling that informs ICF target design and experimental interpretation.
However, due to the great cost of experimentally constraining stopping models in the warm dense regime, the stopping models that are used in the ICF community are often instead validated against other models with varying degrees of accuracy and efficiency.

Broadly, the stopping power models that are applied in the WDM regime fall into four categories: (1) highly detailed multi-atom first-principles models~\cite{magyar2016stopping,ding2018ab,white2018time,white2022mixed,hentschel2023improving}, (2) highly efficient average-atom models~\cite{faussurier2010equation,zylstra2015measurement,hentschel2023improving}, (3) models based on variants of the uniform electron gas~\cite{maynard1982energy,maynard1985born,zimmerman1990recent,moldabekov2015ion,moldabekov2020ion,makait2023time}, and (4) classical or semiclassical models~\cite{li1993charged,li2004stopping,brown2005charged,grabowski2013molecular,graziani2012large}.
Type-(2), (3), and (4) models can be efficient enough to tabulate results across the wide range of thermodynamic conditions required by radiation-hydrodynamic codes that support ICF development, or even evaluated inline~\cite{zimmerman1977lasnex}.
Certain type-(3) models are used to generate high-quality reference data and as a proving ground for method development, but their lack of explicit electron-ion interaction limits their ability to capture some important phenomenology in WDM.
Therefore, type-(1) models are most often used to benchmark and calibrate more approximate type-(2) and (4) models \cite{grabowski2020review,hentschel2023improving}.
Given that there are precious few experiments to validate models in the WDM regime, type-(1) models are particularly valuable for quantifying the influence of details that more approximate models lack.
However, type-(1) models incur large computational costs that are aggravated by the large basis sets and supercells that are required to achieve highly converged results \cite{kononov2023trajectory}.

Thus, the state-of-the-art in algorithms for type-(1) models are mean-field methods based on time-dependent density functional theory (TDDFT)~\cite{correa2018calculating}, including Kohn-Sham~\cite{magyar2016stopping} and orbital-free formulations~\cite{white2018time}.
These methods directly evolve the electronic and nuclear dynamics on the same timescale, going beyond the Born-Oppenheimer (BO) approximation but still typically relying on a classical description of the nuclei.
Even treating the electronic dynamics at a mean-field level, the computing campaigns that use type-(1) models to generate benchmark data for other stopping power models in the WDM regime can require hundreds of millions of CPU hours on some of the world's largest supercomputers.
Opportunities to constrain the accuracy of these models are limited by not only the scarcity of experimental data, but also by the notorious difficulty of developing systematically improvable approximations, particularly for the real-time electron dynamics far from equilibrium that are the central focus of type-(1) stopping power calculations.

However, quantum simulation algorithms executed on fault-tolerant quantum computers provide one potential pathway to realizing systematically improvable stopping power calculations, both in the WDM regime and in general.
In this work, we propose a protocol for implementing such a calculation and analyze its resource requirements to establish a baseline estimate for the cost of outperforming classical computers in accuracy.
Recently, a number of other works have examined the quantum resource requirements for simulating materials represented in first or second quantization~\cite{BabbushSpectra, BabbushLow, BabbushContinuum, Su2021, PhysRevResearch.5.013200, rubin2023fault}.  
While it is now possible to efficiently block encode periodic systems in second quantization~\cite{PhysRevResearch.5.013200, rubin2023fault} and discretizations based on localized orbitals are sometimes used in modeling stopping power~\cite{pruneda2007electronic,zeb2012electronic,maliyov2020quantitative}, they are especially poorly suited to WDM because of the unusual atomic configurations typical to extremes of pressure and temperature. 
Plane-wave representations that are used in first-quantized simulations of materials~\cite{BabbushContinuum,Su2021} provide an efficient representation in which the number of qubits scales logarithmically with the total number of plane waves, and nuclei can be treated on the same footing as electronic degrees of freedom.  
Furthermore, plane-wave calculations can be more directly compared to state-of-the-art classical stopping power calculations based on plane-wave TDDFT.  
Thus, to assess and quantify the prospect of realizing quantum advantages in these simulations we provide constant factor resource estimates of stopping power calculations in first quantization where the projectile is treated quantum mechanically.

There are a variety of ways to compute stopping power in the first-quantized plane-wave representation. 
We focus on sampling the projectile kinetic energy as a function of distance traveled and argue that within the desired accuracy range that naive Monte Carlo sampling at a number of points along the projectile trajectory is more efficient than alternative mean-estimation algorithms that enjoy Heisenberg scaling~\cite{kothari2023mean}. 
After fully accounting for sampling costs, two types of time-evolution protocols, and observable accuracy requirements, we report that full \textit{ab initio} modeling of stopping power for an alpha-particle projectile in deuterium would require $10^{15}-10^{17}$ Toffoli gates, depending on the time-evolution algorithm used, and $10^{3}$ logical qubits at system sizes that are converged to the thermodynamic limit. 
Relaxing the convergence restriction would lower these costs quadratically in the particle number and potentially serve as a WDM benchmark system.  
While there have been a number of works that quantify quantum resource estimates for ground state preparation and sampling from the Hamiltonian eigenbasis~\cite{Lee2020, vonBurg2020, rubin2023fault, Berry2019QF}, this work is among the first to consider a dynamics problem of this scale and real-world significance. 
While the reported resource estimates are high, there are a number of avenues for reducing these. 

\subsection{Stopping power from first principles}
\begin{figure}
    \centering
    \includegraphics[scale=0.85]{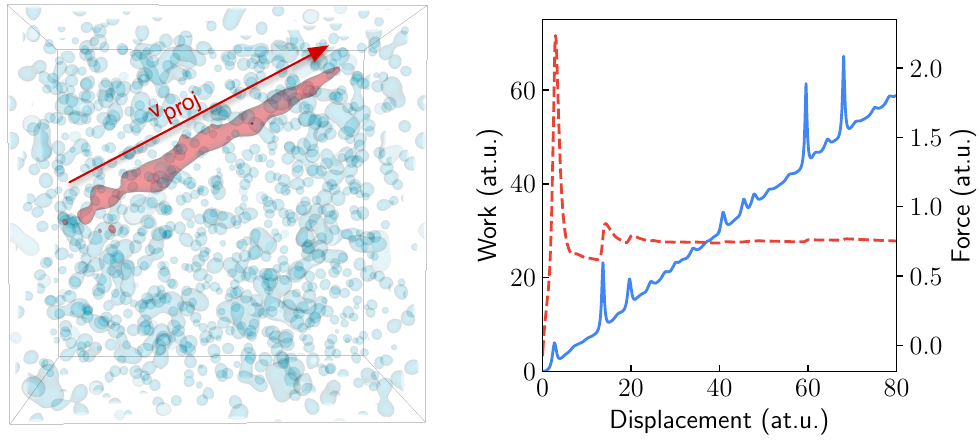}
    \caption{
    (Left) First-principles stopping power calculations involve time evolving a projectile (red) passing through a target medium (blue) while monitoring observables related to energy transfer between them.
    The initial velocity, $v_{\proj}$, is chosen to mitigate trajectory sampling and finite-size error using techniques from Ref.~\cite{kononov2023trajectory}.
    The coupled electron-projectile dynamics are time evolved subject to this initial condition and the work or average force on the projectile is calculated throughout the trajectory.
    (Right) The stopping power is related to the slope of the work that the target does on the projectile as a function of its displacement from its original position (solid).
    A moving average for this slope (dashed) illustrates the rate at which the stopping power estimate converges.
    Close collisions involve large impulses in the work that are essential to capture on average. 
    However, if these relatively rare events are included in the sample, they can dominate the variance for sample-efficient estimates.
    }
    \label{fig:overview_figure}
\end{figure}

While there are many different methods for calculating stopping powers, the most direct method involves time evolving the target and projectile from an initial condition in which the target is stationary and the projectile has some imposed velocity $v_{\rm proj}=k_{\rm proj}/M_{\rm proj}$, rest mass $M_{\rm proj}$, and charge $\zeta_{\rm proj}$.
The stopping power is related to the average rate of energy transfer between the target and projectile over the course of this evolution.
Exemplary results for a classical algorithm that implements this method using Kohn-Sham TDDFT are illustrated in Fig.~\ref{fig:overview_figure}.

The information required to compute a stopping power can be extracted from a number of different observables: the force that the projectile applies to the target, the force that the target applies to the projectile, the work done by the projectile on the target, and the work done by the target on the projectile.
In classical first-principles calculations, the differences among the computational costs of evaluating any of these observables are negligible relative to the overall cost of time evolving the system.
Thus there is no reason to prefer any particular observable, and it is even straightforward to verify the consistency among these quantities (i.e., Newton's third law and the relationship between force and work).
However, the costs of estimating these observables using a quantum algorithm differ significantly.
For shot-noise limited estimation the overall cost will scale with the variance of the observable, whereas Heisenberg-scaling estimation can have costs that scale, in some cases, with the norm of the observable of interest.
Thus we should prefer low-weight observables (i.e., those supported on the projectile quantum register rather than the target) with well-behaved spectra (i.e., energies rather than forces, which exhibit pathologies when estimated naively~\cite{assaraf2000computing,chiesa2005accurate}). 

Another distinction between classical and quantum algorithms for computing stopping powers is the relative cost of time-dependent Hamiltonian simulation.
In classical stopping calculations that use Ehrenfest dynamics with TDDFT, it is common practice to explicitly break energy conservation by maintaining a fixed definite projectile velocity over the course of the evolution.
While energy-conserving calculations in which the projectile is allowed to slow down under the influence of the target will produce equivalent stopping powers, minor technical advantages related to ease of implementation make the fixed velocity approach preferable in practice.
The computational cost of either of these approaches is practically the same for classical simulations, 
but not for quantum algorithms. 
In fact, we will show that within the energy-conserving approach, promoting the projectile to a dynamical and quantum degree of freedom incurs relatively little overhead while facilitating the use of much simpler time-independent quantum simulation algorithms.

\section{Quantum algorithmic protocol for stopping power}
To circumvent the need for time-dependent Hamiltonian simulation, our protocol identifies the dynamical degrees of freedom as the electrons and the single projectile nucleus evolving in the fixed Coulomb field of the remaining nuclei comprising a representative supercell of the target material.
For the projectile velocities at which electronic stopping is most relevant (on the order of 1 atomic unit) ignoring the motion of the target nuclei is justifiable unless their thermal velocities are comparable to the projectile velocity, at temperatures well beyond the WDM regime.
The relevant Hamiltonian, $H$, is
\begin{equation}
    H = H_0 - \frac{\nabla^2_{\proj}}{2 M_{\proj}} - \sum_{i=1}^\eta \frac{\zeta_\mathrm{proj}}{\|r_i - R_\mathrm{proj}\|} + \sum_{\ell=1}^{L}\frac{\zeta_{\proj}\zeta_{\ell}}{\|R_{\ell} - R_{\proj}\|},\label{eq:full_non-BO}
\end{equation}
where $H_0$ is the usual first-quantized BO Hamiltonian for $\eta$ electrons and $L$ classical nuclei~\cite{BabbushContinuum,su2021fault}
and the final three terms account for the projectile nucleus.
The first additional term is the projectile kinetic energy, the second term is the electron-projectile interaction, and the third is the interaction between the quantum projectile and classical nuclei. 

Our protocol consists of four steps: initial state preparation, time evolution, measurement, and postprocessing.
The initial state of the electronic subsystem is drawn from a thermal distribution modeled as a fermionic Gaussian state, while the initial state of the projectile is a Gaussian wave packet in momentum space with a mean velocity corresponding to the projectile velocity and a variance chosen to balance accuracy and efficiency (see Section~\ref{sec:state_preparation}).
From that initial state, the coupled electronic-projectile system is time evolved under $H$ using qubitization (see Section~\ref{sec:time_evolution_with_qubitization}) or Trotterization (see Section~\ref{sec:time_evolution_using_trotterization}).
Analysis and attendant constant-factor resource estimates are presented for both types of time-evolution algorithms.
The measurement step consists of estimating the kinetic energy of the projectile via direct measurement of the kinetic energy operator on the projectile register.
Since we are free to choose a relatively small variance in the nuclear wave packet, we show that shot-noise limited estimation is more efficient than Heisenberg-scaling approaches~\cite{kothari2023mean} unless high accuracy is required (see Section~\ref{sec:estimating_samples}).
From estimates of the instantaneous projectile kinetic energy at a series of distinct times, we can estimate the stopping power using classical postprocessing similar to Fig.~\ref{fig:overview_figure}.
While it is possible to directly mirror the TDDFT strategy on a fault-tolerant quantum computer \--- estimating the projectile energy loss at each of the thousands of time steps along its trajectory \--- to ensure a resource optimal protocol we propose a postprocessing strategy that requires many fewer samples than are typically used in postprocessing TDDFT data (see Section~\ref{sec:estimating_samples}).
Finally, we estimate resource requirements for implementing this protocol to calculate electronic stopping powers for projectile/target pairings relevant to ICF applications in the WDM regime (see Section~\ref{sec:resource_estimates}).

\subsection{Preliminaries}
\label{sec:preliminaries}

We consider a system comprised of $\eta$ electrons, a quantum projectile with mass $M_{\rm proj}$ and charge $\zeta_{\rm proj}$, and $L$ classical nuclei with charges $\zeta_{\ell}$ and positions $R_{\ell}$ in a cubic supercell of volume $\Omega$.
In a plane-wave basis, the associated first-quantized BO Hamiltonian is 
\begin{align}
    H &= T_{\rm elec} + T_{\proj} + U_{\rm elec} + U_{\proj} + V_{\rm elec} + V_{\proj} \label{eq:electron-projectile_hamiltonian} \\ 
    T_{\rm elec} &= \sum_{i=1}^{\eta}\sum_{p \in G}\frac{\|k_{p}\|^{2}}{2}|p\rangle\langle p|_{i} \\
    T_\proj &= \sum_{p\in \tilde{G}} \frac{\left \| k_p - k_{\rm proj}\right\|^2}{2 \, M_{\proj}} \ket{p}\!\bra{p}_{\proj} \label{eq:central_kinetic_term_proj}\\
    U_{\rm elec} &= -\frac{4\pi}{\Omega}\sum_{\ell=1}^{L}\sum_{i=1}^{\eta}\sum_{\substack{p,q \in G\\p\neq q}} \left(\zeta_{\ell}\frac{e^{ik_{q - p}\cdot R_{\ell}}}{\|k_{q-p}\|^{2}} \right)|p\rangle \langle q|_{i}\\
    U_\proj  &= \frac{4 \pi}{\Omega} \sum_{\ell=1}^{L}\sum_{\substack{p,q\in \tilde{G} \\ p\neq q}}\bigg(\zeta_\ell\zeta_{\proj}\frac{e^{ik_{q-p}\cdot R_\ell}}{\norm{k_{p-q}}^2}\bigg)\ket{p}\!\bra{q}_{\proj} \\
    V_{\rm elec} &= \frac{2\pi}{\Omega} \sum_{i\neq j}^{\eta} \sum_{p, q \in G} \sum_{\substack{\nu \in G_{0} \\ (p + \nu) \in G \\ (p - \nu) \in G }} \frac{1}{\|k_{\nu}\|^{2}} |p + \nu\rangle \langle p|_{i} |q - \nu\rangle \langle q|_{j} \\
	V_{\rm proj} &= -\frac{4 \pi}{\Omega} \sum_{i=1}^{\eta}\sum_{\substack{p \in G \\ q \in \tilde{G}}} \sum_{\substack{\nu\in G_0\\(p+\nu)\in G \\ (q - \nu) \in \tilde{G}}}\frac{\zeta_\proj}{\left\| k_{\nu}\right\|^2} |p + \nu\rangle\langle p|_{i} |q - \nu\rangle \langle q|_{\proj} 
\end{align}
where $k_{p} = 2\pi p/\Omega^{1/3}$ represents a plane-wave vector with $p \in G$ and $G = \left[ -(N^{1/3} - 1)/2, (N^{1/3} - 1)/2 \right]^{\otimes 3} \in \mathbb{Z}^{3}$.
$N$ is the size of the plane-wave basis, and we assume a cubic simulation cell of volume $\Omega$. 
Additionally, the set 
\begin{align}
G_{0} = \left[ -(N^{1/3} - 1), (N^{1/3} - 1) \right]^{\otimes 3} \in \mathbb{Z}^{3} \slash \{0,0,0\} \nonumber
\end{align}
is the set of possible differences in momentum (note the range is twice that of the $p$ range in each direction). 
The sets $\tilde{G}$ and $\tilde{G}_{0}$ are defined analogously for the set of plane-wave momenta available to the projectile, taken relative to its initial mean momentum $k_\proj$ with $k_\proj \,\Omega^{1/3}/(2\pi) \in \mathbb{Z}^3$ for compatibility with the supercell's periodicity.
The non-BO contributions due to the projectile are consistent with Eq.~\eqref{eq:full_non-BO}.

In the quantum algorithm the state of the electrons are represented as three signed integers requiring $n_{p} = \lceil\log(N^{1/3})\rceil + 1$ qubits each. The electronic degrees of freedom require a total of $3 \eta n_{p}$ qubits to represent. 
Representing a localized projectile with a sharply peaked momentum distribution in a plane-wave basis requires a large energy cutoff, depending on the variance of the wave packet.
Given a number of plane waves for the projectile $N_{n}$, the number of qubits needed to represent each of the the $xyz$-coordinates is $n_{n} = \lceil \log(N_{n}^{1/3})\rceil +1$ as a set of signed integers, similar to the electronic degrees of freedom.  Thus the total number of qubits to represent the non-BO system is $3\eta n_{p} + 3n_{n}$. We discuss in Section~\ref{sec:grid_resolution} that $n_{n} \approx 3 + n_{p}$ for the target/projectile combinations explored in this work.

\subsection{Initial state preparation \label{sec:state_preparation}}
The WDM regime is typified by an electronic temperature comparable to the electronic chemical potential~\cite{graziani2014frontiers,dornheim2018uniform}, thus it is particularly important to capture the impact of temperature on the initial state.
Ideally we would initialize the system in a product state between the electrons and projectile, in which the electrons are in their exact thermal equilibrium state and the projectile is in a state with a sharply defined velocity, $\rho=\exp(-\beta H_0)/\text{Tr}\left[\exp(-\beta H_0)\right] \otimes |\psi_{\proj}(t=0)\rangle \langle \psi_{\proj}(t=0)|$.

Unfortunately preparing the thermal ensemble on the electronic subsystem is believed to be exponentially hard for generic local Hamiltonians, even on a quantum computer.
Here we assume that the initial thermal distribution of the system is well described by a finite-temperature mean-field approach such as thermal Hartree--Fock or Mermin Kohn-Sham DFT~\cite{kohn1965self,mermin1965thermal}, which is easy to prepare on a quantum computer~\cite{babbush2023meanfield}.
Previous path integral Monte Carlo calculations of the static properties of WDM~\cite{militzer2009path,driver2012all,militzer2021first} suggest that Mermin Kohn-Sham DFT is often an excellent approximation where the domains of feasibility overlap, and it is by far the most popular \textit{ab-initio} approach used in the field. 
The initial state of the calculation is thus chosen to be a Slater determinant
drawn from the solution of a Mermin Kohn-Sham DFT calculation, noting that we do not expect that quantum resource estimates will be sensitive to the choice of exchange-correlation functional.
The use of a thermal Hartree--Fock reference state may also be appropriate, consistent with its use as the conventional starting point for building wave function methods.
The liquid-like ordering common in WDM along with the long-range excitations occurring in stopping power simulations often necessitate 
supercells large enough that a single-point reciprocal-space quadrature is justifiable (e.g., the Baldereschi mean-value point~\cite{baldereschi1973mean} or $\Gamma$ point).
We assume our initial state to have been drawn from such a single-point calculation.

To prepare a sample from this ensemble we use the improved Slater determinant state preparation protocol described in Ref.~\cite{babbush2023meanfield}.
The initial orbitals
\begin{align}
|\psi_{i}\rangle = \sum_{p\in G}c_{i,p}|p\rangle
\end{align}
are used to generate the Slater determinant
\begin{align}
|\psi_{SD}\rangle = | \psi_{1} \wedge \ldots \wedge \psi_{\eta}\rangle,
\end{align}
where $p$ indexes the wavenumbers in a plane-wave basis and $\wedge$ is the antisymmetric tensor product.
This state is prepared with $\tilde{O}(N\eta)$ cost using the
Givens rotation protocol, which is applied sequentially to a second-quantized
representation of the initial state.
This cost is a negligible additive contribution to the time-evolution cost and not accounted for in the resource estimates in Section~\ref{sec:resource_estimates}.

However, $|\psi_{SD}\rangle$ is only a single sample from the initial mean-field density matrix associated with the Mermin Kohn-Sham solution. 
Each Mermin Kohn-Sham orbital $|\psi_i\rangle$ has a temperature-dependent occupation according to the Fermi-Dirac distribution, and thanks to the mean-field nature of the reference state, the probability associated with a particular Slater determinant is proportional to 
the product of probabilities that the corresponding set of single-particle orbitals is occupied (and its complement is unoccupied).
We are careful to note that we are able to circumvent the prohibitive growth in the number of partially occupied Mermin Kohn-Sham orbitals with temperature, which limits the feasible system sizes and temperatures in many classical mean-field approaches to WDM.
This is because any given Slater determinant has support on only $\eta$ orbitals, though we will still need a classical representation of thermally occupied high-energy orbitals.
The state preparation, time evolution, and measurement steps must be repeated $N_s$ times to sample a thermal distribution over initial electronic conditions, and a (potentially) different Slater determinant will be prepared for each initial sample.
This is accounted for in the total sampling requirements and it contributes multiplicatively to the total resource estimates.
Numerical tests indicate that sampling from the attendant canonical and grand canonical ensembles have similar overheads, so we choose to develop our protocol for the canonical ensemble to avoid the need for preparing states with different particle number.
Beyond numerical tests, we generally expect this to be a good choice for the WDM regime due to its low compressibility.

The projectile state $|\psi_{\proj}(t=0)\rangle$ describes the quantum projectile nucleus starting at the same position as in the classical Mermin Kohn-Sham state used to intialize the electronic subsystem, but moving with a velocity $v_{\proj}$.
Preparing the projectile register involves two steps: (1) replacing the corresponding point charge with a Gaussian charge distribution with real-space standard deviation $\sigma_{r}=\sigma_{k}^{-1}$ and (2) translating the average momentum of that charge distribution to the initial momentum of the projectile, $k_\proj$.
In momentum space the resulting initial condition on the projectile wave packet
is \begin{equation} \psi_\proj(k,t=0) =
\sqrt{\frac{1}{(2\pi)^{3/2}\sigma_{k}^3}} \; e^{i k \cdot R_{\rm
proj}} \, e^{-\|k-k_{\proj}\|^2/4\sigma_{k}^2}.\label{eq:init_nuclear_wavepacket} \end{equation}
There is no temperature dependence in the initial nuclear wave packet because it is far from equilibrium in a state with a relatively sharply peaked velocity ($\sigma_{k} \ll k_{\proj}$).
Subsequent non-BO dynamics of the coupled electron-projectile system will
cause this wave packet to slow down such that its mean velocity will decay
linearly, on average, with the average deceleration proportional to the stopping power.
This initial Gaussian wave packet can be prepared using a method in Ref.~\cite{PRXQuantum.3.020364} that contributes an additive $\mathcal{O}(\log N_n)$ cost that is negligible relative to the cost of time evolution.

\subsubsection{Choosing the projectile's initial variance \label{sec:sigma_k}}
\label{subsubsec:choosing_nuclear_variance}

The projectile wave packet will have support on a range of momenta and will disperse (i.e., stop at different rates) if that range is too large.
As long as the momentum gradient of the stopping power is relatively small over the dominant momentum components of the wave packet, this dispersion will be negligible over the relevant time evolution and we can treat the standard deviation as approximately fixed to facilitate resource estimation.
Thus the value of $\sigma_{k}$, a free parameter, should be set to facilitate efficient sampling in the nuclear momentum (computational) basis to realize the fewest circuit repetitions and shortest run time.

While it might appear that we can make the sampling problem arbitrarily efficient by reducing $\sigma_{k}$, there is a trade-off in replacing the nuclear point charge in the BO problem with an explicit wave packet in the non-BO problem.
Making the wave packet too narrow in momentum space will spread out the nuclear charge to an unphysical extent in real space, such that the resulting response will no longer represent a physical nuclear projectile traversing the medium.
One way to understand this is in terms of the equivalent electrostatic potential of the projectile wave packet, 
\begin{equation}
    \frac{\zeta_\proj\,\text{erf}\left(\|r-R_\proj\|/\sqrt{2}\sigma_{r}\right)}{\|r-R_\proj\|},
\end{equation}
where the error function approaches 1 for $\|r-R_{\proj}\| \gg \sigma_{r}$, and the potential appears to be equivalent to that of a point charge.
However, for relatively small values of $\|r-R_{\rm proj}\|$ the effective nuclear charge is reduced along with the strength of the interaction between the projectile and both the electronic and nuclear degrees of freedom.
One might then be tempted to set $\sigma_{r}$ to be consistent with the physical extent of the projectile nucleus (e.g., $\sim 10^{-5}\,a_0$ for a proton or alpha particle, where $a_0$ is the Bohr radius), but the corresponding $\sigma_{k}$ would then be $\sim 10^{5}$ a.u. and the sampling efficiency would be vastly degraded.

To quantify the error associated with replacing the point-like BO projectile nucleus with an explicit non-BO degree of freedom with a finite $\sigma_{k}$ we can consider the difference between the BO electron-projectile force and the non-BO electron-projectile force projected onto a fixed form for the wave packet,
\begin{equation} \Delta S_{e} = -\frac{4 \pi}{\Omega} \sum_{i=1}^{\eta}\sum_{\substack{p,q\in G\\(p-q)\in G_0}} \frac{\zeta_\proj(ik_{p-q}\cdot\hat{v}_{\rm proj})}{\|k_{p-q}\|^2}\ket{p}\!\bra{q}_i\left(e^{ik_{q-p}\cdot R_{\rm proj}}-\sum_{\substack{\nu \in G\\\nu-(p-q)\in G}} \psi_{\rm proj}^*(k_\nu-k_{p-q})\psi_\proj(k_\nu)\right) \label{eq:work_error_operator}.
\end{equation}
It is straightforward to see that this difference vanishes for a projectile wave packet of the form in Eq.~\eqref{eq:init_nuclear_wavepacket} in the $\sigma_{k}\rightarrow \infty$ limit.
While we are using non-BO dynamics to simulate the stopping process, we aim to design an initial projectile state that remains a good approximation to a point nucleus throughout the subsequent dynamics.
This is because we are primarily using non-BO dynamics to avoid the overheads of simulation with an explicitly time-dependent BO Hamiltonian and to allow the electrons to be excited by the nucleus, even while the projectile remains essentially classical in response.

One can think about the difference in Eq.~\eqref{eq:work_error_operator} as quantifying a particular bias in the simulation.
Ideally, we would be able to bound this difference to relate $\sigma_{k}$ to the error that this introduces in the estimate of the stopping power.
However, numerical tests suggest that convenient analytical bounds are too loose to be useful and we instead turn to classical TDDFT calculations for guidance.
In many plane-wave TDDFT calculations the electrostatic potential of the point-like projectile ($\sigma_{r}\rightarrow 0$) is included in the total Hartree potential, which is itself represented in terms of a plane-wave basis set and (not quite) dual real-space grid.
Thus these calculations are themselves subject to a similar source of error in so far as the real-space grid only approximately captures the point-like nature of the projectile, even in the absence of pseudization.
We expect that setting $\sigma_{r}$ to be less than or equal to the real-space grid spacing ensuring a particular degree of convergence in TDDFT forces (typically an order of magnitude or more below the target precision in the stopping power estimate) will introduce a bias that is consistent with the degree of convergence. 

We estimate the size of this bias from TDDFT calculations implemented using an extension of the Vienna \emph{ab initio} simulation package (\textsc{VASP}) \cite{kresse1996efficient,kresse1996efficiency,kresse1999from} described and applied in Refs.~\cite{baczewski2016x,magyar2016stopping,hentschel2023improving,kononov2023trajectory}.
Proton stopping power calculations for a deuterium plasma at a density of 10 g/cm$^3$ and temperature of 1 eV were analyzed for plane-wave cutoffs ranging from 500 eV to 5000 eV and a velocity of 1 a.u.
A cutoff of 1000 eV suffices to converge the estimated stopping power to within $7 \times 10^{-3}$ a.u.\ of the 5000 eV calculation, and the former cutoff corresponds to a real-space grid spacing of $1.75 \times 10^{-1}\,a_0$.
Thus we expect that $\sigma_k \approx 5.7$ a.u.\ suffices to achieve a comparable bias in the force.
A plane-wave cutoff of 4000 eV corresponds to $\sigma_k \approx 11.4$ a.u.\ and reduces this estimated bias by almost two orders of magnitude, and thus we expect $\sigma_{k}=5-10$ a.u.\ to be a good rule of thumb for low-Z projectiles/targets and densities between 1 and 10 g/cm$^3$, relevant to WDM conditions that occur in ICF targets on their way to ignition.
We note that we expect these estimates to be somewhat pessimistic because the supporting TDDFT calculations with different cutoffs each start from a different initial electronic state and contain additional convergence errors from different discretizations of the electronic system. 

\subsubsection{Grid resolution for the projectile wave packet}\label{sec:grid_resolution}

Given a value for $\sigma_k$, we next need to determine the number of plane waves required to 
resolve the kinetic energy of the projectile to within a specified accuracy.
The larger the value of $\sigma_k$, the more plane waves will be required.
We can numerically estimate the number of plane waves by computing the kinetic
energy of the wave packet as a function of the kinetic energy cutoff. 
Taking our initial wave packet as
\begin{equation}
    |\psi_\mathrm{proj}\rangle  = \frac{1}{\sqrt{\mathcal{N}}}
        \sum_{p \in \tilde{G}} e^{-\frac{\left\|k_p\right\|^2}{4\sigma_k^2}} |p\rangle_\proj,
\end{equation}
where the normalization factor is
$\mathcal{N} = \sum_{p \in \tilde{G}} e^{-\frac{\|k_p\|^2}{2\sigma_k^2}}$,
then the kinetic energy of the projectile is given by
\begin{equation}
    \langle T_\proj \rangle =  \frac{1}{\mathcal{N}}\sum_{p \in \tilde{G}}
    \frac{\left\|k_p - k_\proj\right\|^2}{2M_\proj} e^{-\frac{\left\|k_p \right\|^2}{2\sigma_k^2}}\label{eq:ke_proj_sum}.
\end{equation}

Another concern with representing a continuous Gaussian wave packet on a grid is the discretization error which should vanish as $\Omega\rightarrow\infty$.
We can monitor both of these convergence issues by computing the error in the kinetic energy
\begin{equation}
\epsilon_T = \frac{1}{\mathcal{N}}\sum_{p\in\tilde{G}} \frac{\left\|k_p\right\|^2}{2M_\proj} e^{-\frac{\left\|k_p\right\|^2}{2\sigma_k^2}}
    -\frac{1}{\mathcal{N}_\infty} \int dk^3
    \frac{\left\|k\right\|^2}{2M_\proj} e^{-\frac{\left\|k \right\|^2}{2\sigma_k^2}},
    \label{eq:sum_int_err}
\end{equation}
where 
the $\left\|k_\proj\right\|^2$ and $k_p \cdot k_\proj$ terms contributing to the $\left\|k - k_\proj\right\|^2/(2M_\proj)$ factors cancel and vanish, respectively,
and $\mathcal{N}_\infty = (2\pi)^{3/2}\sigma_k^3$.
The integral expression comprising the second term is proportional to the second moment of the normal distribution and evaluates to $3\sigma_k^2/(2M_\proj)$.

\begin{figure}[h!]
    \centering
    \includegraphics[width=8.5cm]{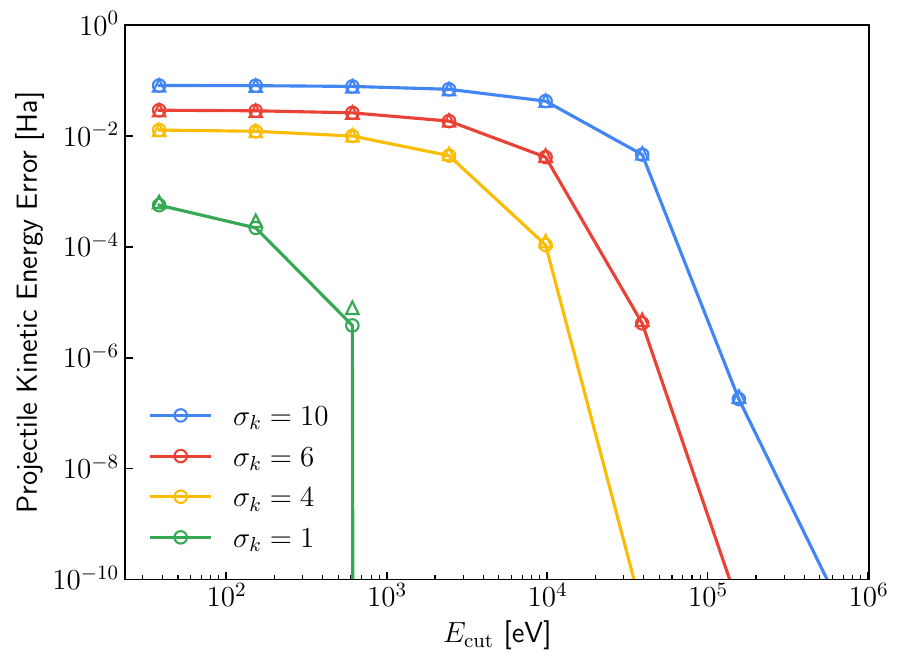}
    \caption{Kinetic energy convergence for a proton projectile with respect to the plane-wave cutoff for different values of the Gaussian wave packet standard deviation $\sigma_k$ in a.u.
    Circles indicate errors evaluated according to \cref{eq:sum_int_err}. Triangles indicate errors calculated by replacing the sum in \cref{eq:sum_int_err} with a truncated integral, demonstrating that the discretization error for the values of $\sigma_k$ chosen in this work is very small.
    }
    \label{fig:gaussian_ke_vs_cutoff}
\end{figure}

For the box sizes considered here ($\Omega^{1/3} \approx 15\,a_0$) and $\sigma_k > 1$ a.u.\ as required by \cref{sec:sigma_k}, we find that the number of plane waves required to achieve a low $\epsilon_T$ can substantially exceed the number that suffices for convergence in classical TDDFT simulations.
\cref{fig:gaussian_ke_vs_cutoff} shows that obtaining a kinetic energy error below $10^{-3}$ Ha with $\sigma_k = 10$ a.u.\ for a proton projectile requires a plane-wave cutoff of approximately $10^5$ eV, 100 times greater than the TDDFT cutoff described in \cref{sec:sigma_k}.
This cutoff corresponds to $N\approx 1.7\times 10^7$ plane waves and would require $n_n = 9$ bits to store each component of the projectile's momenta.
In \cref{tab:quantum_resources_gaussian} we list the resources required for different choices of $\sigma_k$. 

\begin{table}[H]
    \centering
    \begin{tabular}{|ccccc|}
        \hline
        $\sigma_k$ [a.u.] & $E_\mathrm{cut}$ [eV] & $N_n$ & $n_n$ & $\epsilon_T$ [Ha]\\
        \hline
      \hline
1 & $3.8 \times 10^{1}$ & $6.4 \times 10^{1}$ & 3 & $5.6 \times 10^{-4}$ \\
4 & $9.8 \times 10^{3}$ & $2.6 \times 10^{5}$ & 7 & $1.1 \times 10^{-4}$ \\
6 & $3.9 \times 10^{4}$ & $2.1 \times 10^{6}$ & 8 & $4.2 \times 10^{-6}$ \\
10 & $1.6 \times 10^{5}$ & $1.7 \times 10^{7}$ & 9 & $1.8 \times 10^{-7}$ \\
        \hline
    \end{tabular}
    \caption{Number of plane waves $N_n$ and qubits $n_n$ needed to converge the kinetic energy of the projectile wave packet (in 3D) to $\epsilon_T < 10^{-3}$ Ha for different values of the standard deviation $\sigma_k$ and cutoff energy $E_\mathrm{cut}$. Here we assumed a simulation volume of $\Omega^{1/3}=15 \, a_0$.}
    \label{tab:quantum_resources_gaussian}
\end{table}

\subsection{Time evolution using qubitization}\label{sec:time_evolution_with_qubitization}
The first protocol for time-evolution that we consider is using quantum signal processing (QSP) to synthesize the time-independent Hamiltonian propagator~\cite{low2017optimal}, which relies on qubitization~\cite{Low2016}. 
The qubitization walk operator is defined as 
\begin{align}
Q = (2|0\rangle \langle 0 | - I) \cdot \mathrm{PREP}_{H}^{\dagger} \cdot \mathrm{SEL}_{H} \cdot \mathrm{PREP}_{H},
\end{align}
where $\mathrm{PREP}_{H}$ (\textsc{PREPARE}) and $\mathrm{SEL}_{H}$ (\textsc{SELECT}) are defined as
\begin{align}
\mathrm{PREP}_{H}|0\rangle = \sum_{\ell}\sqrt{\frac{\alpha_{\ell}}{\lambda}}|\ell\rangle,\qquad  \mathrm{SEL}_{H}|\ell\rangle = |\ell\rangle \otimes H_{\ell} , \qquad \lambda_H = \sum_{l}\alpha_{l}
\end{align}
and we express $H$ as a linear combination of unitaries (LCU)
\begin{align}
H = \sum_{\ell}\alpha_{\ell}H_{\ell}, \qquad  H_{\ell}^{\dagger}H_{\ell} = \mathbb{I}.
\end{align}
While many works use $Q$ directly as a phase estimation target it was originally shown by Low \textit{et al.}~\cite{low2017optimal} that one can realize the propagator for $H$ with error $\epsilon$ for duration $t$ with 
\begin{align}
\mathcal{O}\left( \lambda t + \frac{\log(1/\epsilon)}{\log\log(1/\epsilon)} \right)
\end{align}
queries to $Q$. Specifically, the query costs are approximately
\begin{align}
2 \left(\lambda t + \frac{3^{2/3}}{2}\left(\lambda t \right)^{1/3} \log^{2/3}(1/\epsilon)\right),
\end{align}
where the expression in the large parentheses is for the polynomial order of the function approximation for $e^{-i(H/\lambda)(\lambda t)}$. This cost is determined by bounding the error in the Jacobi-Anger expansion of the propagator~\cite{BabbushPRA19}.  

An alternative strategy for constructing the propagator discussed in Ref.~\cite{Su2021} and Ref~\cite{Low2018} is simulation based on the interaction-picture algorithm. 
As a proxy for which strategy will be more efficient we can refer to the `parameter region of advantage' plot in Fig.~2 of Ref.~\cite{Su2021}. This plot quantifies for fixed grid spacing $\Delta = \Omega^{1/3}/N^{1/3}$ and particle number $\eta$ which algorithm is more efficient for performing phase estimation. The interaction picture results in lower Toffoli counts for very high resolution simulations with $\Delta < 0.01 \, a_0$.
The qubitization approach outperforms the interaction-picture approach for moderate grid resolutions ($\Delta > 0.01 \, a_0$) with 10 or more electrons.  In this work we consider moderate grid resolutions with a large number of electrons and thus consider the quantum resources necessary to implement QSP.

\subsubsection{Block encoding modifications}
Here we detail the block encoding modifications necessary to account for the non-BO projectile.  
In order to block encode the projectile Hamiltonian we must define
${\prep}_{T_{\rm{proj}}}$, {\prep}$_{U_{\rm{proj}}}$, {\prep}$_{V_{\rm{proj}}}$, {\sel}$_{T_{\rm{proj}}}$, {\sel}$_{U_{\rm{proj}}}$, {\sel}$_{V_{\rm{proj}}}$
such that
\begin{align}
\langle 0| \mathrm{PREP}_{T_{\rm{proj}}}^{\dagger}\mathrm{SEL}_{T_{\rm{proj}}} \mathrm{PREP}_{T_{\rm{proj}}}|0\rangle &= T_{\rm{proj}} / \lambda_{T}^{\rm{proj}} \\
\langle 0| \mathrm{PREP}_{U_{\rm{proj}}}^{\dagger}\mathrm{SEL}_{U_{\rm{proj}}} \mathrm{PREP}_{U_{\rm{proj}}}|0\rangle &= U_{\rm{proj}} / \lambda_{U}^{\rm{proj}} \\
\langle 0| \mathrm{PREP}_{V_{\rm{proj}}}^{\dagger}\mathrm{SEL}_{V_{\rm{proj}}} \mathrm{PREP}_{V_{\rm{proj}}}|0\rangle &= V_{\rm{proj}} / \lambda_{V}^{\rm{proj}} .
\end{align}

The similarity of the Hamiltonian operators for the projectile and the electrons means that much of the same infrastructure for block encoding the electronic Hamiltonian can be used for the electron-projectile Hamiltonian in Eq.~\eqref{eq:electron-projectile_hamiltonian}. The computation of $\lambda$ for each term is thus also \emph{identical} to that in Ref.~\cite{Su2021}, except with $\eta$ replaced with $\eta+\zeta_{\proj}$ and $\lambda_{\zeta}$ corresponds to the total charge of all nuclei treated classically. The $\lambda$ values associated with each additional term are
\begin{align}
 \lambda_\nu^{\proj} &= \sum_{\nu\in \widetilde{G}_{0}} \frac 1{\norm{\nu}^2}, \, \;\;  \lambda_\nu = \sum_{\nu\in G_{0}} \frac 1{\norm{\nu}^2} \,  \\
 \lambda_U^{\proj} &= \frac{\zeta_{\proj} \lambda_\zeta}{\pi\Omega^{1/3}}\lambda_\nu^{\proj}\, \\
 \lambda_{V}^{\proj} &= \frac {\eta\zeta_{\proj}}{\pi \Omega^{1/3}}\lambda_\nu \,\\
 \lambda_{T}^{\rm proj} &= \frac{6\pi^{2}}{M_{\rm proj} \Omega^{2/3}}2^{2(n_{n} - 1)} \\
 \lambda_T^{\rm mean} &= \frac{2\pi \sum_{w\in\{x,y,z\}}|k^w_{\rm proj}|}{M_{\proj}\Omega^{1/3}} \frac{2^{2(n_p-1)}}{2^{(n_p-1)}-1}
\end{align}
where the $\lambda_{T}^{\rm mean}$ corresponds to the cross term obtained from simulating the nuclear kinetic energy in the central momentum frame (Eq.~\eqref{eq:kmean_nuclei_appendix} in Appendix~\ref{app:accounting_nucleus_pw}) and $\lambda_{T}^{\proj}$ corresponds to the LCU 1-norm for the projectile kinetic energy without a shift. A full justification for the projectile $\lambda$ values (including the electronic $\lambda$ values) can be found in Appendix~\ref{app:accounting_nucleus_pw}.
While these $\lambda$ terms are substantial we find them orders of magnitude lower than the electronic kinetic energy $\lambda$ and electronic potential $\lambda$ values.

A considerable difference in the method for block encoding the Hamiltonian for the projectile is in how we account for the success probability of state preparation associated with the potential terms.
In the case without the projectile, the state preparation is ideally split into preparation for the kinetic energy component of the Hamiltonian ($\mathrm{PREP}_{T_\elec}$) and the potential components of the Hamiltonian ($\mathrm{PREP}_{U_\elec + V_\elec}$) as
\begin{align}\label{eq:idealized_prep}
\left(\sqrt{\frac{\lambda_{T}^\elec}{\lambda_{T}^\elec + \lambda_{U}^\elec + \lambda_{V}^\elec}} |0 \rangle + \sqrt{\frac{\lambda_{U}^\elec + \lambda_{V}^\elec}{\lambda_{T}^\elec + \lambda_{U}^\elec + \lambda_{V}^\elec}} |1 \rangle\right)_{a} \otimes \left(\mathrm{PREP}_{T_\elec}|0\rangle\right)_{b} \otimes \left(\mathrm{PREP}_{U_\elec+V_\elec}|0\rangle\right)_{c} \, .
\end{align}
Here, the ancilla qubit $a$ is used to select between the kinetic and potential terms.
The expression for state preparation in Eq.~\eqref{eq:idealized_prep} is not quite what is implemented, due to the fact that the most computationally efficient methods of implementing $\mathrm{PREP}_{U_\elec+V_\elec}$ give a normalization factor of approximately $4\left(\lambda_{U}^\elec + \lambda_{V}^\elec\right)$.
The factor of approximately $1/4$ corresponds to the probability of success in the state preparation, so the actual state preparation is
\begin{align}
\widetilde{\mathrm{PREP}}_{U_\elec+V_\elec}|0\rangle_{f} \otimes \mathbb{I} = \frac{1}{2}|0\rangle_{f} \otimes \mathrm{PREP}_{U_\elec+V_\elec} + \frac{\sqrt{3}}{2}|1\rangle_{f} \otimes \mathrm{PREP}^{\perp}_{U_\elec+V_\elec}.
\end{align}
The ancilla qubit $f$ flags successful state preparation (on the $|0\rangle$ state) for potential terms.
The actual amplitude (which needs to be squared for the probability) is not exactly $1/2$.
The expression to determine the probability of success, $p_{\nu}$, is given in Eq.~(106) of Ref.~\cite{Su2021}.
To account for the imperfect probability of success, the choice of whether to apply the kinetic or potential term is based on both the ancilla $a$ and the success flag $f$.
A further subtlety is that we test $i\ne j$ in the block encoding of $V_\elec$, with a further failure in the case where $i=j$ (representing self-interaction).
The total electronic $\lambda$ was thus defined
\begin{equation}\label{eq:lamnonamp}
    \lambda_H^\elec = \max\left[ \lambda_T^{\elec}+ \lambda_U^{\elec}+\lambda_V^{\elec}, [\lambda_U^{\elec}+\lambda_V^{\elec}/(1-1/\eta)]/p_\nu \right] ,
\end{equation}
where the factor of $1-1/\eta$ accounts for testing $i\ne j$.

We also previously considered the case where amplitude amplification is performed for the state preparation over $G_{0}$.
Typically the value of $p_\nu$ is slightly smaller than $1/4$, so a single step of amplitude amplification can be used to give a new probability of success (see Eq.~(117) of Ref.~\cite{Su2021})
\begin{equation}
    p_{\nu}^{\rm amp} = \sin^2\left[ 3\arcsin (\sqrt{p_\nu})\right]\, .
\end{equation}
This probability will be slightly less than 1.
A further step of amplitude amplification could be performed, but the extra cost would more than outweigh any advantage from the boosted success probability.
With this boosted success probability one can otherwise take exactly the same approach, so Eq.~\eqref{eq:lamnonamp} would be changed to
\begin{equation}
    \lambda_H^\elec = \max\left[ \lambda_T^{\elec}+ \lambda_U^{\elec}+\lambda_V^{\elec}, [\lambda_U^{\elec}+\lambda_V^{\elec}/(1-1/\eta)]/p_\nu^{\rm amp} \right] .
\end{equation}
The only amendment is in the probability.

For the current application where we simulate the electrons and the projectile we now have three kinetic energy terms ($T_{\rm elec}$, $T_{\rm proj}$, and $T_{\rm mean}$) along with four potential terms ($U_\elec$, $U_\proj$, $V_\elec$, and $V_\proj$).  Of the additional two potential terms the $U_{\rm proj}$ term requires a new state preparation over $\widetilde{G}_{0}$ while $V_{\rm proj}$ is incorporated into the original state preparation from $\widetilde{\mathrm{PREP}}_{U_\elec+V_\elec}$ with slightly different logic for \textsc{SEL}$_{\elec}$ (discussed more in more detail in Appendix~\ref{app:accounting_nucleus_pw}). Just as before we consider the tradeoffs in costs for state-preparation of the potential terms and adjusting the success probabilities for imperfect preparations.  
Because the number of qubits required to represent the projectile is expected to be larger than the electrons (see Section~\ref{sec:grid_resolution} for more details) the respective $\nu$ superpositions can be prepared on the same register with additional controls accounting for the fact that $n_{p} < n_{n}$.
The new $\lambda$ value accounting for preparing the total state over both electronic and projectile degrees of freedom is thus
\begin{align}
\lambda_H = \max\left[ \lambda_T^{\elec} +\lambda^{\proj}_T+\lambda^{\rm mean}_T+ \lambda_U^{\elec} +\lambda^\proj_U+\lambda_V^{\elec} +\lambda^\proj_V, [\lambda_U^{\elec} +\lambda_V^{\elec}/(1-1/\eta)+\lambda^\proj_V]/p_\nu +\lambda_U^\proj/p_{\nu,\proj} \right] .
\end{align}
When using amplitude amplification $\lambda_H$ has an identical expression except $p_{\nu,\proj}$ and $p_{\nu}$ are exchanged for $p_{\nu,\proj}^{\rm{amp}}$ and $p_{\nu}^{\rm amp}$ 
where ${\rm amp}$ superscript corresponds to the probability of success when performing amplitude amplification for the respective state preparations.

In the electronic-only case, additional registers for selecting between $U_{\elec}$ and $V_{\elec}$ along with kinetic and potential terms were needed.
With the additional terms we need additional registers to control the application of each operator.  The adjustment of the probabilities is detailed in Appendix~\ref{selstateprep}.
We now provide a summary accounting of the block encoding costs and relegate the detailed derivation to Appendix~\ref{app:accounting_nucleus_pw}.
For consistency with Ref.~\cite{Su2021}, we consider the costs of each line in Table II of that work.
In the following list, the numbers correspond to the lines of that table.
\begin{enumerate}\label{en:costs}
    \item[C1.] The cost of the preparation of the registers selecting between the components of the Hamiltonian is changed to $6n_T-1$, where $n_T$ is the number of qubits used in inequality tests.
    That is the cost for the six inequality tests in the preparation of the three registers used, and accounting for one of those inequality tests using one fewer qubit (see Appendix \ref{selstateprep}).
    We are also assuming that the temporary ancillas are retained so these inequality tests may be inverted without Toffolis.
    There will be another Toffoli for the AND or OR in the preparation of the qubit flagging whether the kinetic or potential energy is applied.
    There are another two Toffolis for controlled swaps for selecting the qubit that selects between electron and projectile components.
    That gives a total of $6n_T+2$.
    \item[C2.] The cost of preparing the superposition over $\eta$ values of $i,j$ is unchanged from that in Ref.~\cite{Su2021}.
    There it is given as $14n_\eta+8b_r-36$ (where $n_{\eta} = \lceil \log(\eta)\rceil$) for preparing a superposition and flagging $i\ne j$.
    \item[C3.] The state preparation of the $w,r,s$ registers for the kinetic energy operators will have cost $2(2n_n+9)+2(n_n-n_p)+20$. The first amendment over the $2(2n_p+9)$ expression in Ref.~\cite{Su2021} is to replace $n_p$ with $n_n$ because we need to account for the larger basis for the projectile. The second amendment is to include a $2(n_n-n_p)$ term to account for the extra control of the Hadamards between the electron and projectile parts.
    The third amendment is to include the cost of the inequality test for preparing a second $w$ register (20 Toffolis) and the controlled swaps (4 Toffolis).
    \item[C4.] The cost of swaps into the working registers is $12\eta n_p+6n_n+4\eta-6$.
    Because we are now selecting between the $\eta$ electron registers and the projectile register for $i$ but \emph{not} $j$, we need another two Toffolis, which changes $-8$ to $-6$.
    The extra $6n_n$ term is for the controlled swaps in the case when we are selecting the projectile register for $i$.
    There is a factor of $6$ rather than $12$ because it is only for $i$, not $i$ and $j$ as for the electron registers.
    \item[C5.] The select cost of the kinetic energy terms is increased from $5(n_p-1)+2$ to $5(n_n-1)+2$ due to the need to account for the larger number of qubits for the projectile.
    There is another Toffoli for the selection between the square and the product of the momentum with the momentum offset for the projectile, for a total of $5n_n-2$.
    \item[C6.] For the cost of preparing the $1/\|\nu\|$ state, we first need to replace $n_p$ with $n_n$.
    The other amendment is that we need to introduce a cost of $n_n-n_p+1$ for the extra controls on the Hadamards.
    There are $n_n-n_p-1$ double controls, where the double control may be applied with a Toffoli that can be inverted with Clifford gates.
    There is also an extra singly-controlled Hadamard, which needs another Toffoli for inversion.
    That gives the extra cost $(n_n-n_p-1)+2=n_n-n_p+1$.
    That gives a total cost of $3n_n^2+16n_n-n_p-6+4n_{\mathcal{M}}(n_n-1)$ for preparing the $1/\|\nu\|$ state.
    \item[C7.] The QROM cost or $R_\ell$ is unchanged at $\lambda_{\zeta}+{\rm Er}(\lambda_{\zeta})$.
    \item[C8.] For the addition and subtraction cost of $\nu$ into momentum registers, only half of it is $n_p$ replaced with $n_n$, because we are only allowing the projectile momentum in one temporary register.
    That gives a Toffoli cost $12(n_n+n_p)$.
    \item[C9.] For the phasing cost we simply replace $n_p$ with $n_n$ to account for the projectile momentum, to give $6n_n n_R$.
\end{enumerate}

The complexity of the block encoding is still dominated by the controlled swaps (C4).  In all cases we find that the block encoding cost multiplied by $\lambda_H$ computed with boosting the success probability of the potential $\mathrm{PREP}$ using amplitude amplification is smaller than without.
\subsection{Time evolution costs using product formulas}\label{sec:time_evolution_using_trotterization}
The second method we investigate for implementing time evolution uses product formulas to implement the electron-projectile propagator. In this section we consider a real-space grid Hamiltonian model of the electronic structure instead of the full electron-projectile system. We justify this consideration based on the fact that the electronic degrees of freedom are the dominant simulation costs. In order to simulate the time evolution using a product formula, there are three main parts.
\begin{enumerate}
    \item Computing the potential energy in the position basis and applying a phase according to that energy.
    \item Computing the kinetic energy in the momentum basis and applying the corresponding phase.
    \item Performing a QFT between the two bases.
\end{enumerate}

The product formula simulation is performed by alternating steps 1 and 2, using the QFT to switch the basis.
The complexity is expected to be largest for step 1, because this requires computing the potential energy between $\eta(\eta-1)/2$ pairs of electrons, whereas the kinetic energy just requires summing $\eta$ momenta squared.

The main difficulty in calculating the potential energy is in computing the approximation of the inverse square root.
This was addressed in Ref.~\cite{CodyJones2012}, where it was stated that the function could be approximated to within 32 bits within 5 iterations, given a suitably chosen starting value for the iteration.
We have numerically tested this approach, and found that it is only accurate if the starting value is not too far from the correct inverse square root of the argument.
Accuracy of one part in $2^{32}$ is only obtained over a range of less than a factor of 5 for the argument.

To improve on that computation, we consider a hybrid approach combining the QROM function interpolation of Ref.~\cite{Sanders2020} with the Newton-Raphson iteration of Ref.~\cite{CodyJones2012}.
There are a number of variations that one could consider, depending on how the function interpolation is performed and how many steps of Newton-Raphson iteration are used.
We find that excellent performance is obtained by using a single step of QROM interpolation with a cubic polynomial, followed by a single step of Newton-Raphson iteration. A further optimization targeting high-order product formulae is that the Newton-Raphson iteration is generalized to find  $b/\sqrt{x}$ instead of $1/\sqrt{x}$ which saves complexity by avoiding multiplying the potential by the product formulae coefficient.
There is also a choice of how many points are used in the interpolation, and we find that using two points within each factor of 2 of the argument gives relative error within about one part in $4\times 10^8$.
This is almost as high precision as that claimed in Ref.~\cite{CodyJones2012}, and works over the full range of input argument.

To give the costing for this procedure more precisely, we first need to compute the sum of squares of the difference between each of electron's $xyz$-components. This arithmetic has a cost of three squares.
According to Lemma 8 of Ref.~\cite{Su2021}, the complexity is $3n^2-n-1$ when each of the three numbers has $n$ bits.
The resulting number has no more than $2n+2$ bits.
The Toffoli complexity of the QROM to output interpolation parameters is then $4n+2$, using a variable spacing QROM as in Ref.~\cite{Sanders2020}.
As an example, say we were performing the variable spacing QROM on 8 qubits.
The integer ranges that would be used would be
\begin{align}
    0,1,2,3,[4,5],[6,7],[8,11],[12,15],[16,23],[24,31],[32,47],[48,63],[64,95],[96,127],[128,191],[192,255],
\end{align}
where e.g.\ $[16,23]$ is used to indicate the range of integers.

Then for the interpolation, we can describe it by the interpolation for the regions $[1,3/2]$ and $[3/2,2]$.
For example, $[16,23]$ is 16 times a range within $[1,3/2]$, so we can use the parameters chosen for $[1,3/2]$ appropriately scaled for this multiplying factor.
For the region $[1,3/2]$ we can use the interpolation
\begin{equation}
    \frac 1{\sqrt{x}} \approx a_0 - a_1 (x-1) + a_2 (x-1)^2 - a_3 (x-1)^3 \, ,
\end{equation}
with
\begin{align}
    a_0 &= 0.99994132489119882162, \\
    a_1 &= 0.49609891915903542303, \\
    a_2 &= 0.33261112772430493331, \\
    a_3 &= 0.14876762006038398086.
\end{align}
This is followed by a step of Newton-Raphson as
\begin{equation}
    y \mapsto \frac 12 y(3+\delta-y^2 x),
\end{equation}
where $\delta = 5.1642030908180720584\times 10^{-9}$.
These parameters were found by numerically minimising the maximum relative error over the region.
It was then found that the relative error is no more than about $2.5821\times 10^{-9}$ within this interval.
The constants $a_j$ are appropriately scaled for $x$ in the range $[2^m,(3/2)2^m]$ as
$a_0\mapsto a_0/2^{m/2}$, $a_1\mapsto a_1/2^{3m/2}$, $a_2\mapsto a_2/2^{5m/2}$, $a_3\mapsto a_3/2^{7m/2}$, with $x-1$ replaced with $x-2^m$.
(The constant $\delta$ is unchanged.)

For $x$ in the range $[3/2,2]$, one can use 
\begin{equation}
    \frac 1{\sqrt{x}} \approx a_0 - a_1 (x-3/2) + a_2 (x-3/2)^2 - a_3 (x-3/2)^3 \, ,
\end{equation}
with
\begin{align}
    a_0 &= 0.81648515205385221995, \\
    a_1 &= 0.27136515484240234115, \\
    a_2 &= 0.12756148214815175348, \\
    a_3 &= 0.044753028579153842218, \\
 \delta &= 3.6279794522852781448\times 10^{-10}.
\end{align}
These parameters were similarly found by a numerical optimisation to reduce the error, and give relative error no more than about $1.8140\times 10^{-10}$ in this region.

Writing the polynomial in this way makes it appear as if many powers and multiplications are needed.
The computation can be significantly simplified by rewriting it as (for the case of the range $[1,3/2]$)
\begin{equation}
    a_0 - (x-1) \{a_1 - (x-1) [a_2 - a_3 (x-1)]\} = a_0 - a_1 (x-1) + a_2 (x-1)^2 - a_3 (x-1)^3.
\end{equation}
Thus, only three multiplications are needed to compute the polynomial, and the multiplications are the most costly part.
We have also written it with a polynomial of $x-1$ rather than $x$.
This requires one subtraction, but reduces the number of bits needed to represent $x-1$ (versus $x$) by 1, and reduces the number of bits for each of $a_0,a_1,a_2$ by 1.
This subtraction can be performed with Clifford gates because removing a leading 1 from $x$ can be performed with a CNOT gate.
We have a similar form for the polynomial in the range $[3/2,2]$.

It is also found that the initial polynomial interpolation may be given to only 15 bits of precision, and the resulting accuracy of the approximation after the step of Newton-Raphson is still about one part in $10^8$.
For a rough estimate of the complexity of the arithmetic, we can assume it is performed with no more than 15 bits at this step.
As discussed in \cite{Sanders2020}, the complexity of multiplying two real numbers is approximately the square of the number of bits.
This is because less significant bits can be omitted in the calculation.
If we are using 15 bits for each multiplication in the interpolation here, the complexity of the three multiplications is about $3\times 15^2=675$.
This is the dominant cost in the arithmetic, and the subtractions have significantly lower cost.
Three subtractions on 15 bits have Toffoli cost about 45 (computing $x-1$ or $x-3/2$ is not included here because it can be performed with Clifford gates).

For the step of Newton-Raphson, we can estimate the complexity of the square of a 15-bit number as $15^2$ Toffolis.
Then for the other two multiplications, if we aim for, for example, 24 bits of accuracy, then the complexity is $2\times 24^2$.
With another 24 Toffolis for the subtraction, the overall complexity is about 2136 (excluding the complexity of the sum of squares and QROM).
With the number of bits in each direction being $n=6$, the sum of squares has complexity 101, and the QROM has complexity 26 Toffolis.
Those are trivial complexities compared to the other arithmetic for the inverse square root, and would bring the total to about 2263.

There are a couple of additional considerations for the complexity not discussed in the simplified discussion above.
First, for small $x$ the inverse square root is large, so in the multiplication of $x$ by the approximation of $1/\sqrt{x}$ the assumptions in the estimate of the complexity do not hold.
In order to avoid needing to use additional bits of precision in the multiplications to account for that, we can instead use bit shifts.
First, we strip pairs of leading zeros from $x$.
Since $x$ has $2n+2$ bits, the complexity is $n(n+1)$, which is 42 for $n=6$.

In this example, one may remove 2, 4, 6, 8, 10, or 12 leading zeros.
(There cannot be all zeros which would correspond to two electrons at the same location.)
These alternatives would require moving a number of bits which is 14 minus the number of leading zeros.
The Toffoli cost corresponds to the number of bits moved, which gives a total of $12+10+8+6+4+2=42$ Toffolis.
At the end one would need to shift the approximation of the inverse square root back again.
This can be performed on the result of the QROM interpolation before multiplying in the Newton-Raphson iteration.
That would have a Toffoli complexity no more than $15n$ bits giving we are computing the QROM interpolation to 15 bits.
In the example with $n=6$ it is 90 Toffolis.
These two costs are relatively trivial compared to the overall cost of the step, and would bring it to about 2395 Toffolis.
In this cost we have taken the specific example of $n=6$.
The $n$-dependent cost can be given as
\begin{equation}
    2136 + (3n^2 - n - 1) + (4n + 2) + n (n + 1) + 15n = 2137 + 4 n^2 + 19 n \, .
\end{equation}

A further consideration is the need to uncompute the arithmetic.
If we were to compute all the pairwise Coulomb potentials, sum then phase by the sum, we would then need to uncompute the sum.
The number of Toffolis makes it infeasible to uncompute by retaining qubits.
However, we can phase by each individual pairwise Coulomb potential and uncompute the arithmetic with Clifford gates by retaining about 2000 qubits used in the calculation.
This approach would be reasonable given the algorithm is likely to need a large number of logical qubits already.
A difficulty then is that a the potential may need to be multiplied by a constant before phasing.
If we were to compute the complete Coulomb potential before phasing then that complexity would be trivial, but it will be significant if we need to do it for each pairwise potential.

If we were to use just the standard Lie-Trotter product formula then the length of the time step could be chosen such that no multiplication were needed.
Higher-order product formulae would need time steps with irrational ratios, so multiplications would be needed.
However, that can be avoided if we instead compute the factor as part of the QROM interpolation and Newton-Raphson.
If we are aiming to compute $b/\sqrt{x}$, then we can simply multiply the constants in the QROM interpolation by $b$, and replace 3 with $2+b^2$ in the Newton-Raphson step.
That will give the desired factor with the same Toffoli complexity as before.

\subsubsection{Estimating number of Trotter steps}

According to Theorem 4 of Ref.~\cite{Low2022trotter}, for a real-space grid Hamiltonian defined for orbital indices $\{j, k, l, m\}$ and spin indices $\{\sigma, \tau\}$ of the form
\begin{align}\label{eq:ueg}
H = \sum_{j, k,\sigma}\tau_{j,k}a_{j,\sigma}^{\dagger}a_{k,\sigma} + \sum_{l, m,\sigma,\tau}\nu_{l,m}a_{l,\sigma}^{\dagger} a_{l,\sigma}a_{m, \tau}^{\dagger} a_{m, \tau}
\end{align}
the spectral norm error in a fixed particle manifold for an order-$k$ product formula $S_k(t)$ can be estimated as
\begin{equation}\label{eq:Low_bound}
    \norm{S_k(t)-e^{-itH}}_{W_\eta}=\mathcal{O}\left((\norm{\tau}_{1}+\norm{\nu}_{1,[\eta]})^{k-1}\norm{\tau}_1\norm{\nu}_{1,[\eta]}\eta\, t^{k+1}\right).
\end{equation}
Here, $a_{j,\sigma}^\dagger$ and $a_{j,\sigma}$ are creation and annihilation operators, and
the norms are defined as 
\begin{align}
\|\tau\|_{1} &= \max_{j} \sum_{k}|\tau_{j,k}| \\
\|\nu\|_{1,\left[\eta\right]} &= \max_{j} \max_{k_{1} < ... < k_{\eta}} \left(|v_{j,k_{1}}| + ... + |v_{j, k_{\eta}}| \right).
\end{align}
If the constant of proportionality is $\xi$, then breaking longer evolution time $t$ into $r$ intervals gives error
\begin{equation}\label{eq:constant_factor_trotter_scaling}
    \approx \xi (\norm{\tau}_{1}+\norm{\nu}_{1,[\eta]})^{k-1}\norm{\tau}_1\norm{\nu}_{1,[\eta]}\eta\, t^{k+1}/r^k \, .
\end{equation}
In order to provide a simulation to within error $\epsilon$, the number of time steps is then
\begin{equation}
    r \approx t^{1+1/k} (\norm{\tau}_{1}+\norm{\nu}_{1,[\eta]})^{1-1/k}(\xi \norm{\tau}_1\norm{\nu}_{1,[\eta]}\eta/\epsilon)^{1/k}.
\end{equation}

\begin{figure}[H]
    \centering
    \includegraphics[width=8cm]{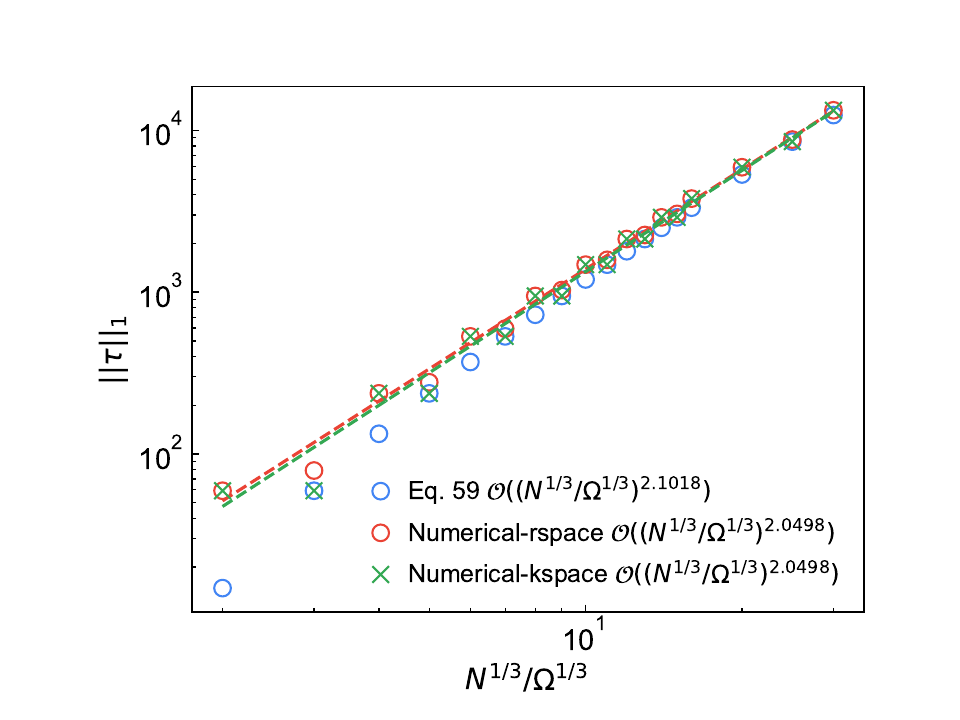}
    \includegraphics[width=8cm]{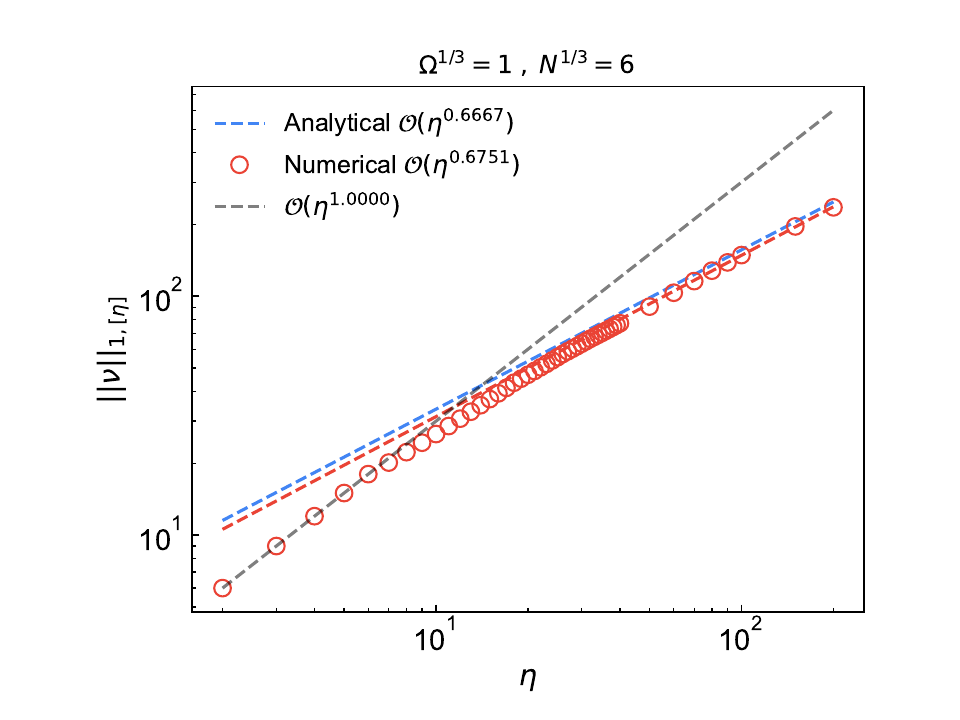}
    \caption{Comparison between the analytical asymptotic value of the the norms $\|\tau\|_{1}$ and $\|\nu\|_{1,\left[\eta\right]}$.  For a given reciprocal lattice sampling defined as  $k_{\nu} = \frac{2\pi \nu}{\Omega^{1/3}}, \;\; \nu \in G, \;\; G = \{ \frac{-N^{1/3} - 1}{2}, \frac{N^{1/3} - 1}{2} \}$, the real space grid is defined as $r_{p} = \frac{p \Omega^{1/3}}{N^{1/3}}, \;\; p \in G$ satisfying shifted (centered) discrete Fourier transform $|r_{p} \rangle = \frac{1}{\sqrt{N}}\sum_{\nu} e^{i k_{\nu} \cdot r_{p}}|k_{\nu}\rangle$ and thus $k_{\nu}\cdot r_{p} = \frac{2\pi}{N^{1/3}}\nu\cdot p$.  The grey dashed line for the $\|\nu\|_{1,\left[\eta\right]}$ norm plot describes the scaling of the rightmost four red (numerical) points.  The red dashed line corresponds to a fit to the rightmost five red (numerical points).  Similarly in the $\|\tau\|_{1}$ norm plot the $N$ scalings are determined by fitting the rightmost five points. }
    \label{fig:tau_nu_scaling}
\end{figure}

Our goal is to numerically determine the constant factor $\xi$ for a high-order product formula. We first provide the constant factors of the norms used in the error scaling Eq.~\eqref{eq:Low_bound} and describe their convergence to asymptotic values. The scaling of the norms is
\begin{align}
    \norm{\tau}_{1} &= \mathcal{O} \left( \frac{N^{2/3}}{\Omega^{2/3}} \right) \\
    \norm{\nu}_{1,[\eta]} &= \mathcal{O} \left( \frac{\eta^{2/3} N^{1/3}}{\Omega^{1/3}} \right) \, .
\end{align}
To find the constant factor for $\norm{\tau}_{1}$, note that it corresponds to the kinetic energy of a single electron, and is
\begin{equation}\label{eq:tau_norm_db}
    \norm{\tau}_{1} = \max_p \frac{\|k_p\|^2}2 = \max_p \frac{4\pi^2\|p\|^2}{2\Omega^{2/3}} = \frac{4\pi^2}{2\Omega^{2/3}} 3 [(N^{1/3}-1)/2]^2 \approx \frac{3\pi^2N^{2/3}}{2\Omega^{2/3}}.
\end{equation}
The norm $\norm{\nu}_{1,[\eta]}$ comes from Eq.~(K4) of \cite{Su2021} which uses the the potential operator 
\begin{equation}
    V = \frac{N^{1/3}}{2\Omega^{1/3}}\sum_{i\ne j} \sum_{p,q} \frac 1{\|p-q\|} \ket{p}\!\!\bra{p}_i \ket{q}\!\!\bra{q}_j .
\end{equation}
and corresponds to the potential energy for a single electron with the other electrons packed around it as closely as possible. For $p,q$ where there is a unit grid spacing, the volume is $\eta \approx (4/3)\pi R^3$, giving a radius of $R\approx [(3/4)\eta/\pi]^{1/3}$.
The potential energy is the integral of $1/r$ over a sphere from 0 to $R$.
That gives
\begin{equation}
    \int_0^R 4\pi r^2/r\, dr = 4\pi \int_0^R r \, dr = 2\pi R^2 = 2\pi [(3/4)\eta/\pi]^{2/3}.
\end{equation}
Thus, $\norm{\nu}_{1,[\eta]}$ including constant factors is approximately
\begin{equation} \label{eq:nu_norm_db}
    \norm{\nu}_{1,[\eta]} \approx \pi^{1/3} (3/4)^{2/3}\frac{\eta^{2/3} N^{1/3}}{\Omega^{1/3}} .
\end{equation}
To explore how quickly the norms converge to their asymptotic values (Eq.~\eqref{eq:nu_norm_db} and Eq.~\eqref{eq:tau_norm_db}) we plot each norm as a function of basis size and number of particles in Figure~\ref{fig:tau_nu_scaling}. We find that $\|\tau\|_{1}$ converges relatively quickly with respect to the grid spacing ($N^{1/3}/\Omega^{1/3}$) but $\|\nu\|_{1,\left[\eta\right]}$ does not converge until approximately 50 particles. 
In order to determine the constant $\xi$ in Eq.~\eqref{eq:constant_factor_trotter_scaling} we numerically determine the spectral norm of Eq.~\eqref{eq:Low_bound} for a variety of product formulas for a variety of systems scaling in $N$ and $\eta$. To avoid building an exponentially large matrix, we adapt the power method to determine the spectral norm of $\Delta(t) = S_k(t)-e^{-itH}$ as the square root of the maximal eigenvalue of $\Delta(-t)\Delta(t)$.  Numerically taking the square root would halve the precision, so we instead use the power iteration to estimate the spectral norm of half an application of $\Delta(-t)\Delta(t)$.  The full algorithm is outlined in Algorithm~\ref{alg:spectral_norm} which is implemented in the Fermionic Quantum Emulator (FQE)~\cite{rubin2021fermionic}. Using the FQE we can target a particular particle number sector, projected spin $s_{z}$ sector, and use fast time evolution routines based on the structure of the Hamiltonian in Eq.~\eqref{eq:ueg}.  Our numerics involved 64 orbital (128 qubit) systems involving 2-4 particles.

\begin{figure}[H]
    \centering
    \includegraphics[width=13cm]{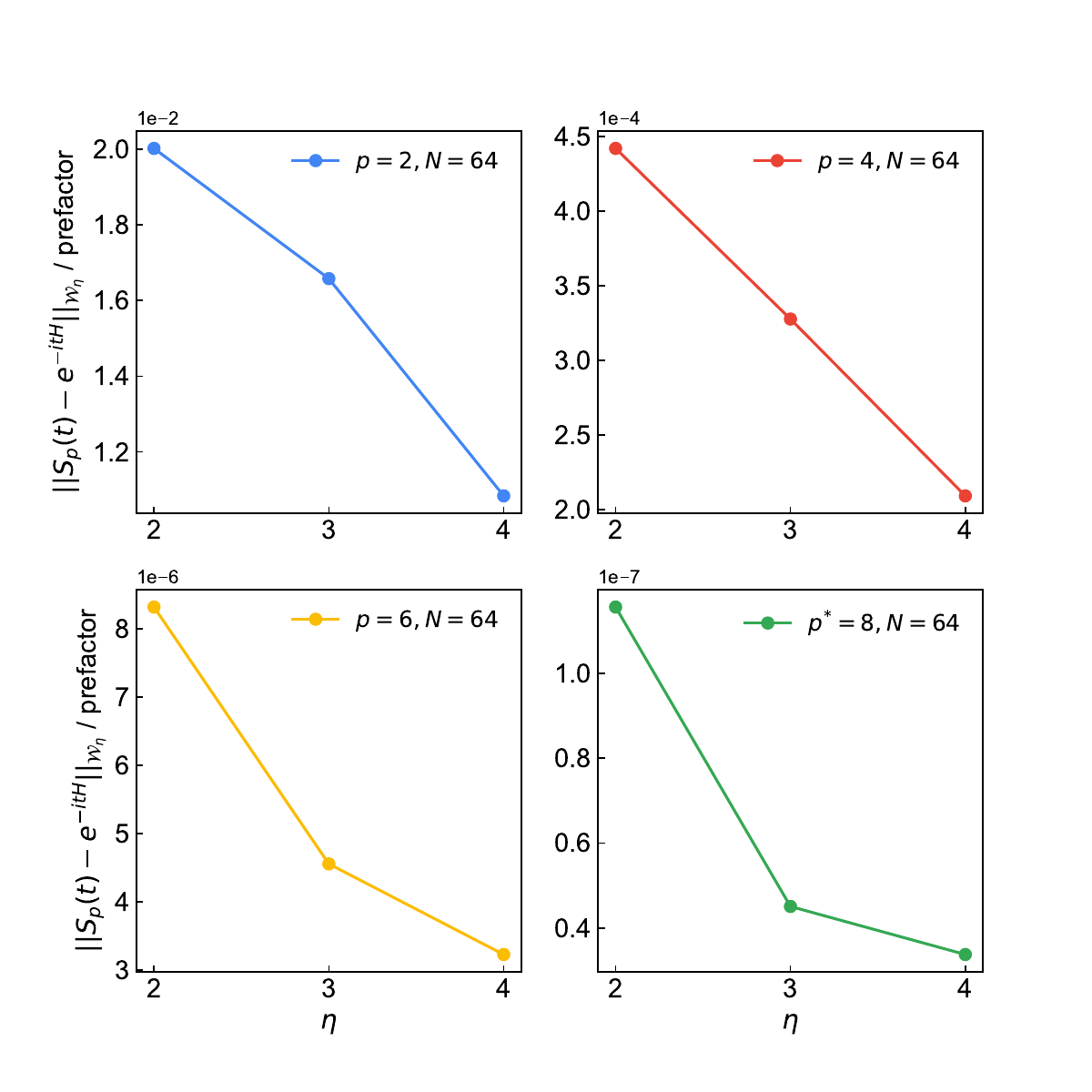}
    \caption{$N=64$, $\Omega=5$ grid based Hamiltonian convergence of the prefactor $\xi$ with respect to particle number. $\xi = 3.4 \times 10^{-8}$ is used as an upper bound for the prefactor in all subsquent product formula resource estimates. $p$ indicates the product formula order (star on the lower right plot indicates a numerically determined product formula). The blue points (upper left) are for the Strang product formula, red points (upper right) are for 4$^{\rm th}$-order Suzuki-Trotter, yellow (lower left) are for 6$^{\rm th}$-order Suzuki-Trotter, and green (lower right) corresponds to a custom 8$^{\rm th}$-order formula described in Appendix~\ref{app:cost_of_trotter}.}
    \label{fig:n64_prefactor_xi}
\end{figure}

In the Figure~\ref{fig:n64_prefactor_xi} we determined $\xi$ by explicitly calculating the spectral norm of the difference between the exact unitary and a bespoke 8$^{\rm th}$-order product formula discribed in Appendix~\ref{app:cost_of_trotter}. The `prefactor' variable corresponds to $(\norm{\tau}_{1}+\norm{\nu}_{1,[\eta]})^{k-1}\norm{\tau}_1\norm{\nu}_{1,[\eta]}\eta\, t^{k+1}$ for $t=0.65$. For $N=64$ $\eta=4$ we estimate a $\xi = 3.4 \times 10^{-8}$ which is the value of the rightmost point.

For the $8^{\rm th}$-order product formula, each step requires 17 exponentials.
Each exponential has a complexity on the order of $2395\eta(\eta-1)/2$.  Combining the constant factors, norm computation, and number of Toffolis required per exponential allowed us to calculate the Toffoli and qubit complexities for time evolution via product formula.  We provide comparative costs to QSP in Section~\ref{sec:resource_estimates}.

 \SetKwComment{Comment}{/* }{ */}
\RestyleAlgo{ruled} 
\begin{algorithm}[H]
\caption{Power iteration algorithm to compute the spectral norm $\|\Delta(t)\|_{\mathcal{W}_{\eta}} = \Gamma$}\label{alg:spectral_norm}
\KwData{$\epsilon$, $\Delta(t) = S_k(t)-e^{-itH}$}
$\psi_{i=0} \gets \frac{1}{\sqrt{|\mathcal{H}|}}\sum_{j}|j\rangle$\;
$\delta \gets \infty$\;
$\Gamma_{-} \gets 0$\;
$\Gamma \gets 0$\;
\While{$\delta \geq \epsilon$}{
  $\psi_{i+\frac{1}{2}} \gets \Delta(t) \psi_{i}$\;
  $\Gamma \gets \|\psi_{i+\frac{1}{2}}\|$\;
  $\psi_{i+1} \gets \Delta(-t) \psi_{i + \frac{1}{2}}$\;
  $\psi_{i+1} \gets \psi_{i+1}/\|\psi_{i+1}\|$\;
  $\delta \gets |\Gamma_{-} - \Gamma|$\;
  $\Gamma_{-} \gets \Gamma$\;
}
\end{algorithm}

\subsection{Projectile kinetic energy estimation}\label{sec:estimating_samples}

Our protocol estimates the stopping power from a time series for the projectile kinetic energy loss over the course of the electron-projectile evolution.
Here we analyze the sampling overheads for two approaches to estimating the projectile kinetic energy, one at the standard quantum limit and the other with Heisenberg scaling.
This allows us to estimate the number of circuit repetitions required to achieve stopping power estimates with a target accuracy, and thus the aggregate Toffoli count for the entire protocol.
For either approach, an estimate of the total number of samples needs to account for both the thermal distribution of the electrons in the initial state, the variance of the kinetic energy operator, and the effect of time evolution on both of these.
Short of implementation and empirical assessment, there is no clean way to precisely bound the number of samples required to compute the stopping power while taking into account all of these factors and we rely on a few simplifying assumptions to facilitate analysis.

\begin{figure}[h!]
    \centering
    \includegraphics[scale=0.5]{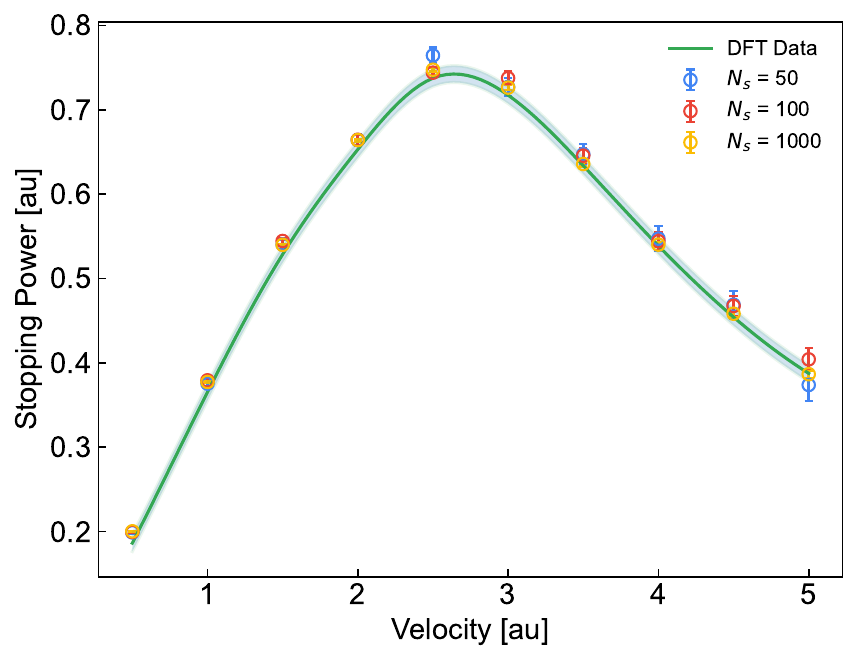}
    \includegraphics[scale=0.5]{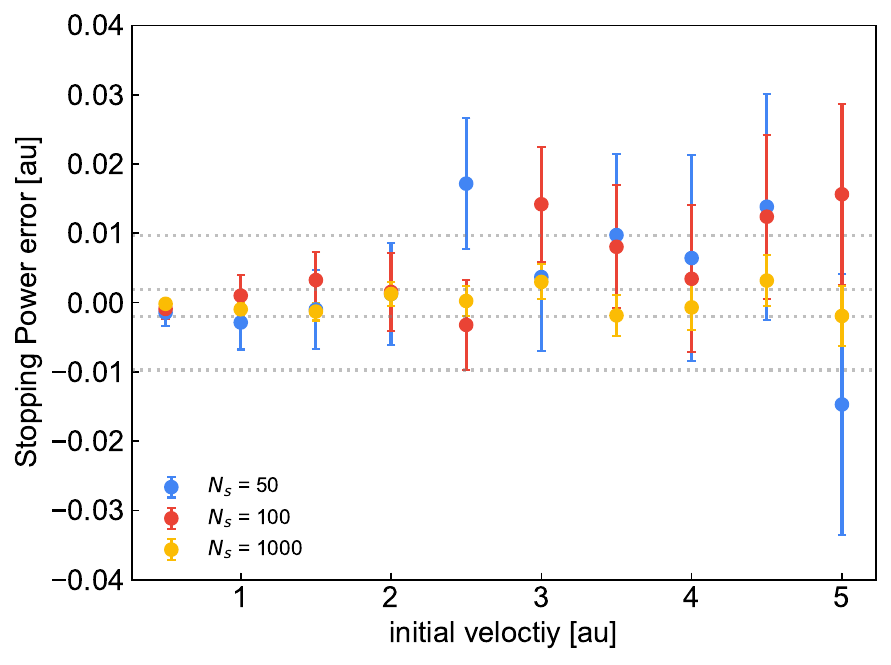}
    \caption{(Left) Comparison between the TDDFT stopping power and the stopping power computed through random sampling of the expected value of a Gaussian wave packet's kinetic energy. The shaded area represents an error of 0.01 a.u.\ demonstrating that the main features of the stopping curve can be resolved with a higher error threshold.
    (Right) Error in stopping power as a function of the projectile's initial velocity and number of samples ($N_s$) used to estimate the Gaussian wave packet's kinetic energy.  Horizontal lines represent an accuracy in the stopping power of 0.1 eV/\AA\ ($\approx$ 0.002 a.u.) and 0.5 eV/\AA\ ($\approx 0.01$ a.u.).}
    \label{fig:gaussian_stopping_sampling}
\end{figure}

Our estimate of the sampling overhead for the first approach (at the standard quantum limit, see Fig.~\ref{fig:gaussian_stopping_sampling}) is based on classical Monte Carlo sampling of the projectile's kinetic energy,
given in \cref{eq:ke_proj_sum}, using the $k_{\proj}(t)$ from a plane-wave TDDFT calculation with classical nuclei and $\sigma_k$ set according to the discussion in Section~\ref{sec:state_preparation}.
The variation in $\sigma_k$ is assumed to be negligible over the timescale associated with electronic stopping, which is short relative to the timescale over which such a wave packet would diffuse thanks to the large difference between the electron and projectile masses.
We estimate the stopping power by computing the slope of the kinetic energy change of the projectile as a function of time.
We take 10 points within the simulated time interval and extract the slope and its error through least-squares regression. In \cref{fig:gaussian_stopping_sampling} we compare the number of samples required to resolve the stopping power to within 0.1 eV/\AA 
$\ \approx 0.002$ a.u. We find that between $10^1$ and $10^3$ samples are required depending on the desired accuracy in the stopping power, with the sample cost growing with the velocity of the projectile.
In practice we expect to require only $10^1-10^2$ samples for accuracy relevant to applications in WDM.
As shown in \cref{app:precision_stopping}, this corresponds to a precision (standard error) in the individual kinetic energy points of approximately 0.1 Ha. 
If the desired accuracy in the stopping power is lowered from 0.1 eV/\AA~ to 0.5 eV/\AA~ (corresponding to the green shaded region in \cref{fig:gaussian_stopping_sampling}) then the number of samples required drops by a factor of 10.

However, this estimate does not directly account for the sampling overhead associated with capturing the thermal distribution of the electrons.
Because the projectile and medium are far from equilibrium and will remain so over the timescale of our simulation, we do not expect that variance of the wave packet to depend on the electronic temperature and these two sources of randomness are independent.
A complete assessment of the associated sampling overhead would require evaluating an ensemble of full quantum dynamics simulations in which the initial states are thermally distributed, thus we leave this to future work.
However, we expect the sampling overhead associated with capturing the thermal distribution of the electrons to already be accounted for in the overhead associated with sampling the final observable, provided we measure a low variance observable like the kinetic energy of the projectile.
We note that we have tested another related strategy to estimate the sampling overhead, involving Monte Carlo estimation of the increase in energy of the electronic system. 
It was found to have a substantially larger variance and required an order of magnitude more samples to obtain comparable precision in the stopping power.

The second method we consider is a Heisenberg-scaling kinetic energy estimate based on an algorithm by Kothari and O'Donnell~\cite{kothari2023mean} (later referred to as KO).  This mean-estimation algorithm has $\mathcal{O}(\sigma^{2}/\epsilon)$ scaling of the time-evolution oracle, a quadratic advantage over Monte Carlo sampling, but requires coherent evolution of a Grover-like iterate constructed from the time-evolution unitary and a phase-oracle constructed from the projectile kinetic energy cost function--\textit{i.e.}\ the projectile kinetic energy \textsc{SELECT}.  To compare this strategy to expectation value estimation by Monte Carlo sampling we estimate the Toffoli complexity of time-evolution based on QSP and determine the constant factors associated with constructing the Grover-like iterate.  Details on the phase-oracle construction are described in Appendix~\ref{app:knockout_algorithm}. Reference~\cite{kothari2023mean} describes a decision problem associated with the mean-estimation task which can be lifted to full expectation value estimation through a series of classical reductions. Using the assumption that we have a fairly accurate estimate of the kinetic energy (valid for an almost classical projectile) the number of calls to the core decision problem is expected to be a small integer multiple. Thus our cost estimates focus on Toffoli counts associated with the core decision problem.

\begin{figure}[H]
    \centering
    \includegraphics[width=8cm]{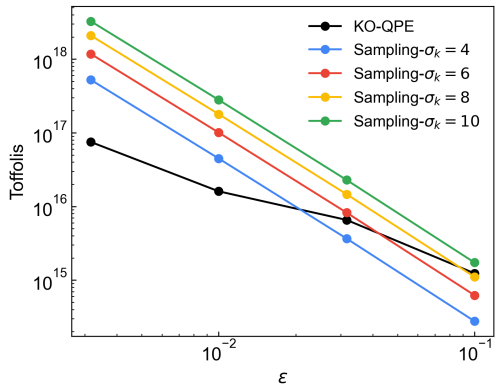}
    \caption{The Toffoli cost of estimating the projectile kinetic energy with traditional Monte Carlo sampling and the mean estimation algorithm from Kothari and O'Donnell~\cite{kothari2023mean} (KO).  Both techniques have standard error that linearly depends on square root of the variance of the observable. The number of samples required for fixed standard error in Standard Monte Carlo mean estimation scales as $\mathcal{O}(1/\epsilon^{2})$ while the KO algorithm scales as $\mathcal{O}(1/\epsilon)$ but with larger constant factors originating from code (circuit) for the random variable and quantum phase estimation on the Grover like iterate used in the algorithm.  More details of the algorithm's main subroutines are provided in Appendix~\ref{app:knockout_algorithm}.  }
    \label{fig:ko_algorithm_comparison}
\end{figure}

To facilitate resource estimation we have built a model of the entire protocol using the Cirq-FT software package~\cite{khattar_tanuj_2023_8161656}, which allows us to quantify the cost of each subroutine.  In Figure~\ref{fig:ko_algorithm_comparison} we plot this estimate (in black) along with estimates for standard Monte Carlo estimates based on the variance of the Gaussian wave packet representing the projectile which assumes no spreading of the particle.  In the low precision regime needed for the stopping power calculations standard sampling has a computational advantage due to the lower overhead (smaller prefactors). Coincidentally, the crossover is just beyond the $\epsilon$ necessary for the kinetic energy observable precision.  In applications where higher precision is necessary (e.g., stopping at/below the Bragg peak or other dynamics problems entirely) the KO algorithm likely provides a computational advantage over sampling despite the classical reduction and phase estimation overheads.

\section{Resource estimates for ICF-relevant systems}\label{sec:resource_estimates}
\begin{figure}
    \centering
    \includegraphics[width=8cm]{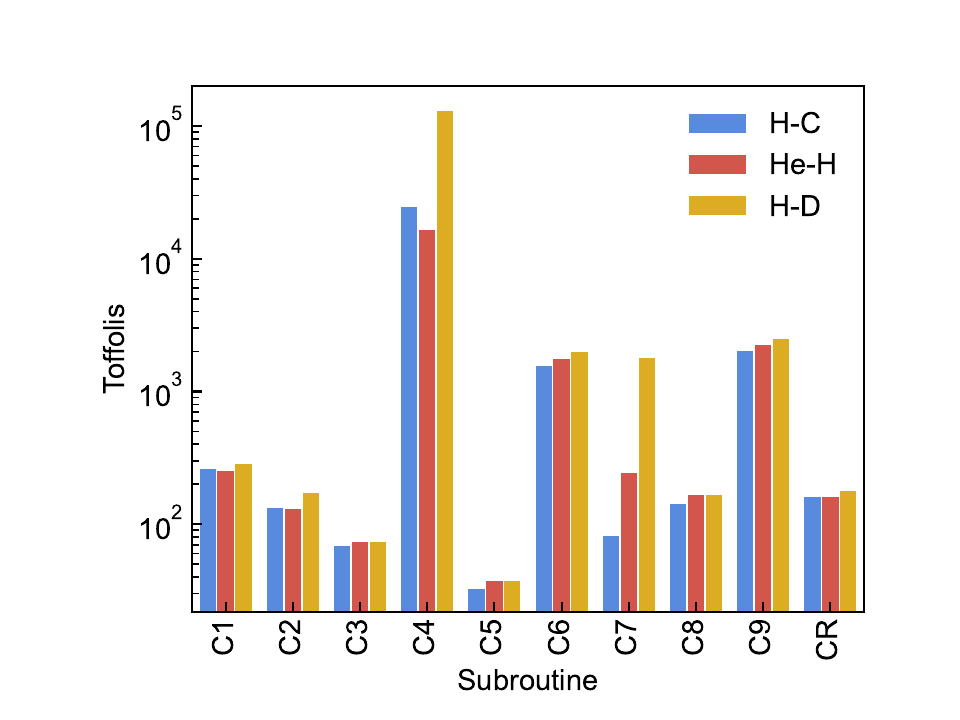}
    \caption{Subroutine costs for each component of implementing the block-encoding.  The labels C$\{n\}$ correspond to the costs enumerated in protocol ~\ref{en:costs}. CR is the reflection cost which given the additional register and augmented $\lambda$ is $n_{T} + 2 n_{\eta} + 6 n_{n} + n_{M} + 16$ where $n_{T} = 10 + \log(\lambda/\epsilon)$, with $\epsilon$ being the target precision of time evolution, and $n_{\eta} = \lceil \log(\eta)\rceil$.}
    \label{fig:block_encoding_cost_breakdown}
\end{figure}

\begin{figure}[H]
    \centering
    \begin{picture}(400,175)
    \put(-50,0){\includegraphics[width=8cm]{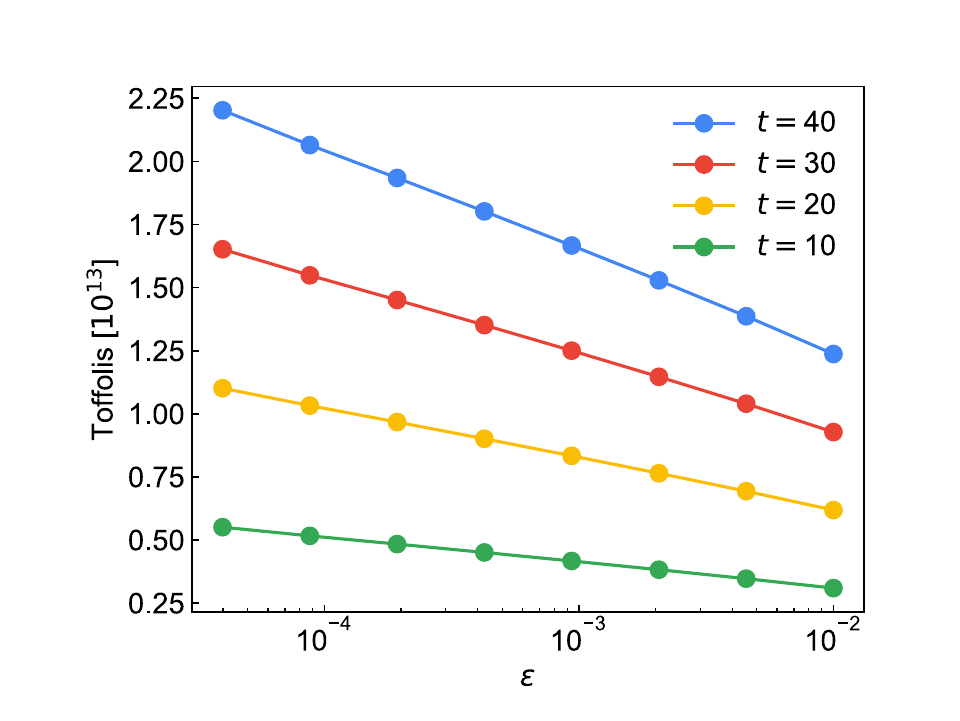}}
    \put(200,0){\includegraphics[width=8cm]{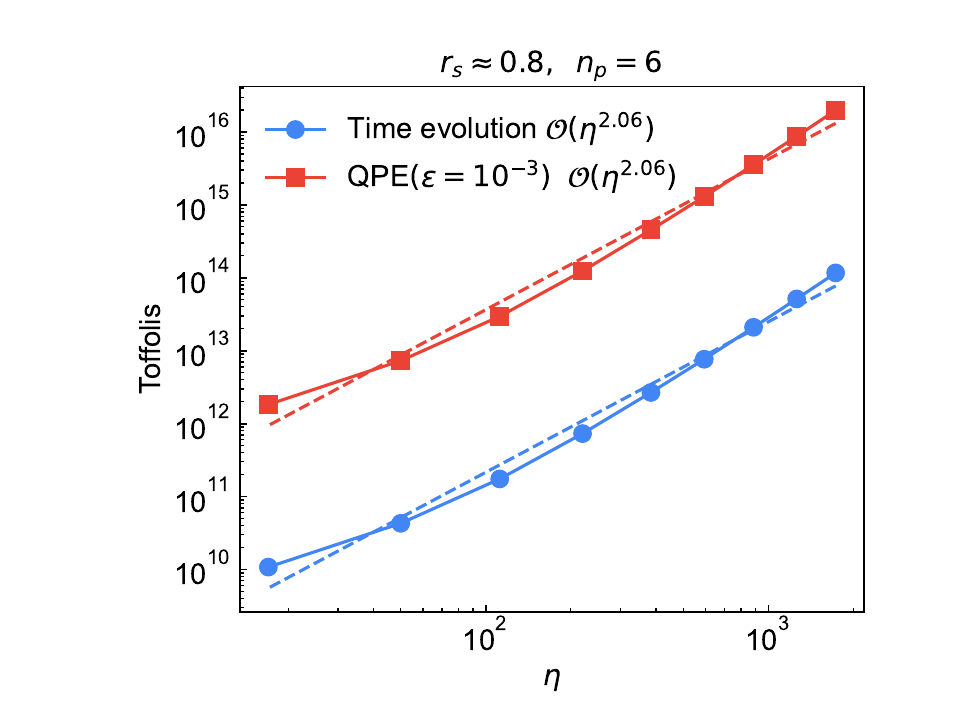}}
    \put(-10,165){(a)}
    \put(240,165){(b)}
    \end{picture} 
    \caption{a) Toffoli complexity to synthesize the system propagator of the Alpha + Hydrogen system for time $t=10,20,30,40$ (in atomic units) for a range of fidelities. Using the lowest sample complexity considered $N_{s}\approx 50-100$ and $10$ total times to estimate the slope the Toffoli complexity is $\approx 200$ times the Toffoli costs shown. b) Toffoli scaling with respect to particle number at fixed $r_{s}$ and a fixed number of planewaves (increasing grid resolution).  Time evolution cost is expected to scale as somewhere in between $\mathcal{O}(\eta^{4/3})$ and $\mathcal{O}(\eta^{8/3})$ using quantum signal processing. In blue the time evolution cost for $t=1$ is shown demonstrating the expected scaling along with constant factors. Constant factors associated with QPE are shown in red for $1/t = \epsilon = 10^{-3}$ which is should proportionally increase the cost. The displayed constant factor resources are in line with what is demonstrated in Ref.~\cite{Su2021} for constant $r_{s}\approx 1$.\label{fig:time_costs_scaling} }
\end{figure}

We present resource estimates for  stopping power calculations in three systems relevant to ongoing efforts aimed at characterizing errors in transport property calculations used in the design and interpretation of ICF and general high-energy density physics experiments~\cite{grabowski2020review}.
This allows us to explore how costs vary with the projectile and target conditions over the relevant phase space.
The first system is an alpha-particle projectile in a hydrogen target at a density of 1 g/cm$^{3}$. 
The second system is a proton projectile in a deuterium target at a density of 10 g/cm$^{3}$. 
The third is a proton projectile in a high-density carbon target at 10 g/cm$^{3}$.
Details of associated Ehrenfest TDDFT calculations of the latter two can be found in Ref.~\cite{hentschel2023improving}.
Each system's relevant classical parameters are defined in Table~\ref{tab:fusion_systems}.
\begin{table}[H]
    \centering
    \begin{tabular}{|cccccccc|}
    \hline 
      Projectile + Host   & Volume [$a_{0}^{3}$] & $\eta$ [electrons] & Wigner-Seitz radius [$a_0$]& $E_{\mathrm{cut}}$ [eV] (Ha) & $N^{1/3}$ & $\Delta$ [$a_{0}$] & $n_{p}$ \\
      \hline \hline
      Alpha + Hydrogen    & 2419.68282 & 218  & 1.383 & 2000 (73.49864) & 53 & 0.25330 & 6 \\
      Proton + Deuterium & 3894.81126 & 1729 & 0.813 & 2000 (73.49864) & 63 & 0.24974 & 6 \\
      Proton + Carbon  & 861.328194 & 391  & 0.807 & 1000 (36.74932) & 27 & 0.35239 & 5 \\
        \hline
    \end{tabular}
    \caption{
    Summary of ICF-relevant systems considered in this work and associated classical simulation parameters.
    $a_{0}$ is the atomic Bohr radius, $E_{\rm cut}$ is the cut off energy used in classical TDDFT calculations to model the system, which corresponds to a grid spacing in one direction of $N^{1/3}$ using a spherical cutoff, $\Delta$ is the grid spacing of the TDDFT calculations, and $n_{p}$ is the number of qubits needed to achieve a similar resolution along one grid dimension.
}
    \label{tab:fusion_systems}
\end{table}
For each of the systems we consider a stopping power calculation with a projectile kinetic energy of 4 a.u.\ and a projectile wave packet variance of $\sigma_{k}=6$ a.u.  This allows us to determine the number of bits of precision needed to represent the projectile wavepacket.  Considering the costs in the previous section we tabulate the block encoding costs $C_{B.E.}$, $\lambda$, and number of logical qubits, for each system in Table~\ref{tab:quantum_resources_fusion_systems}.  For all systems it was considerably cheaper to amplitude amplify the state preparation cost instead of reweighting the kinetic and potential terms. 
\begin{table}[H]
    \centering
    \begin{tabular}{|ccccccc|}
    \hline
      Projectile + Host  &  $M_{\mathrm{proj}}/M_{\mathrm{proj,H}}$ & $n_{n}$ & $k_{\mathrm{proj}}$ @ 4 a.u. & $C_{B.E.}$ [Toffolis] & Num. Qubits & $\lambda$ \\
      \hline \hline
      Alpha + Hydrogen    &  3.9726 & 8 & 29376 & $2.498\times 10^{4}$ & 5650 &   1744784.42 \\
      Proton + Deuterium &  1 & 8 & 7344  & $1.423\times 10^{5}$ & 33038 & 88202784.59\\
      Proton + Carbon  &  1 & 8 & 7344  & $3.836\times 10^{4}$ & 8841 &  7727607.07\\
        \hline
    \end{tabular}
    \caption{
    Summary of quantum algorithmic parameters and costs associated with the systems listed in \cref{tab:fusion_systems}.
    Each column of the table is as follows: System description in terms of the projectile and host type, mass of the projectile $M_{\mathrm{proj}}$ relative to the proton mass $M_{\mathrm{proj,H}}$, the number of bits for the projectile, $n_{n}$, for projectile wave packet variance of $\sigma_{k}=6$ a.u., the cost of block encoding the system, the required number of logical qubits, and total system $\lambda$.}
    \label{tab:quantum_resources_fusion_systems}
\end{table}
To further analyze the cost breakdown and to demonstrate the expected $\tilde{\mathcal{O}}(\eta)$ block encoding complexity we plot the Toffoli requirements for each subroutine outlined in the protocol~\ref{en:costs} in Figure~\ref{fig:block_encoding_cost_breakdown}. As expected controlled swaps of each electron into the working register for performing \textsc{SELECT} (C4) dominates the costs by an order of magnitude or more for each system. It is unlikely that this step can be further improved within this simulation protocol and representation. 
Finally, we can estimate the total Toffoli costs for performing time evolution on the electron-projectile system. In Figure~\ref{fig:time_costs_scaling}a we plot the total Toffoli counts for evolving the Alpha + Hydrogen system for times $t=10,20,30,40$ in units of a.u. for various infidelities $\epsilon$. As the Toffoli complexity scales logarithmically in $\epsilon$ there is little change in the total Toffoli complexity with infidelity. Time linearly scales the query complexity which is already linearly proportional to $\lambda$.  Given the $\lambda$ values in Table~\ref{tab:quantum_resources_fusion_systems} and the block encoding costs the $10^{13}$ Toffoli gates for small constant $t$ values is not unexpected. While this is the price of one state preparation at time $t$ it has already been discussed in Section~\ref{sec:estimating_samples} that an additional $N_{s} \approx 50-100$ samples for $10$ points are needed to reach the desired accuracy of the stopping power estimate. To probe costs for smaller systems we examine the cost of systematically shrinking the unit cell at fixed Wigner-Seitz radius. Given the scaling of qubitization, $\tilde{\mathcal{O}}\left( \frac{\eta^{2}}{\Delta^{2}} + \frac{\eta^{3}}{\Delta}\right)$, fixing the number of planewaves and shrinking the unit cell at fixed particle density corresponds to increasing the grid resolution $\Delta$.  Expressing the total complexity in terms of $\eta$, the Toffoli complexity is expected to scale somewhere between $\mathcal{O}(\eta^{4/3})$ and $\mathcal{O}(\eta^{8/3})$ depending on which term is dominant in the qubitization costs. This value is plotted in blue in Figure~\ref{fig:time_costs_scaling}b with a slope of approximately $\mathcal{O}(\eta^{2})$.  For reference we provide the QPE scaling costs assuming the QPE precision in $10^{-3}$ demonstrating that time-evolution with sampling can be substantially cheaper than eigenvalue estimation. While decreasing the system size while maintaining fixed grid resolution is possible we are only able to decrease the number of gridpoints by powers of two. Shrinking the system by powers of two quickly leads to nonphysically realistic system sizes and thus we focus on shrinking the system with increasing grid resolution which leads to a quadratic decrease in complexity with the number of particles at fixed Wigner-Seitz radius.

We now make a comparison of the total Toffoli and logical qubit costs to estimate the stopping power. This requires ten kinetic energy estimations at times from $t=1$ to $t=10$. Each time evolution is constructed with infidelity $\epsilon=0.01$.  A comparison between building the propagator with QSP and the $8^{\rm th}$-order product formula is shown in Table~\ref{tab:total_costs} which includes a factor of $N_{s}=50$ accounting for the sampling overhead at each of the 10 time points. The product formula numerics use a prefactor of $\xi = 3.4 \times 10^{-8}$ and the analytical values for the $\|\tau\|_{1}$ and $\|\nu\|_{1, \eta}$ norms.
\begin{table}
    \centering
    \begin{tabular}{|cccccc|}
    \hline 
      Projectile + Host   & $\eta$ & QSP Toffoli & Product Formula Toffoli & QSP Qubits & Product Formula Qubits \\
      \hline \hline
      Alpha + Hydrogen (50\%)  &  28 & $5.593 \times 10^{14}$ & $1.124 \times 10^{13}$ & 1749 & 2666 \\
       Alpha + Hydrogen (75\%) &  92 & $2.033 \times 10^{16}$ & $3.069 \times 10^{14}$ & 3309 & 3902\\
       Alpha + Hydrogen        & 218 & $1.992 \times 10^{17}$ & $1.399 \times 10^{15}$ & 5650 & 6170 \\
      Proton + Deuterium        &1729 & $2.121 \times 10^{20}$ & $2.079 \times 10^{17}$ & 33038 & 33368\\
      Proton + Carbon         & 391 & $2.225 \times 10^{18}$ & $1.074 \times 10^{16}$ & 8841 & 9284\\
        \hline
    \end{tabular}
    \caption{Comparison of the total Toffoli cost for time-evolution using QSP or product formulas for 10 uniformly spaced times starting from $t=1$ and going to $t=10$ with infidelity $\epsilon = 0.01$ including $50$ samples to measure the kinetic energy of the projectile. The smallest Alpha + Hydrogen system used $n_{p}=5$ while all other systems used $n_{p}=6$.  For all systems the projectile kinetic energy register used $n_{n}=8$ bits. The number of qubits for the product formula is estimated based on the system size plus an upper bound to the number of ancilla needed for performing the polynomial interpolation multiplications and Newton-Raphson step floating point arithmetic. }
    \label{tab:total_costs}
\end{table}
\section{Discussion}
We have described a quantum protocol for estimating stopping power and derived constant factor resource estimates for systems relevant to ICF. To our knowledge this is the first analysis of a practically relevant quantum time dynamics simulation. It is also the first specific proposal for how fault-tolerant quantum computers can contribute to the development of inertial fusion energy platforms. While the overall resource estimates are high, we expect that the product formula estimates are loose and further algorithmic innovations are possible. Supporting this optimism are the orders-of-magnitude improvements in constant factors for second-quantized chemistry simulation seen over the last five years~\cite{Lee2020, vonBurg2020, rubin2023fault, Berry2019QF}.

We estimated that fully converged (in system size and basis-set size, with respect to TDDFT) calculations for an alpha particle projectile stopping in a hydrogen target would require $10^{17}$ Toffoli gates using QSP as the time-evolution routine. If $8^{\rm th}$-order product formulas are used to build the electron-projectile propagator then we estimate that approximately $10^{15}$ Toffoli gates would be required. Both strategies require $10^{3}$ logical qubits to represent the system and a modest number of additional ancilla required for QROM and implementing product formulas. This estimate is for a system that is substantial in size (219 quantum particles) and corresponds to a calculation that has only been classically tractable using mean-field-like methods like TDDFT or more approximate models. Scaling down the system to a benchmark scale (29 quantum particles) would require substantially fewer resources ($10^{13}$ Toffoli gates) and could be used to quantify the accuracy of TDDFT calculations and other approximate dynamics strategies. 

In order to determine the constant factors we compiled all time-evolution subroutines that contribute to the leading-order complexity. To implement time-evolution we considered two methods to construct the electron-projectile propagator: QSP with qubitization and high-order product formulae. For QSP we extended the block encoding construction of Ref.~\cite{Su2021} to account for the non-Born-Oppenheimer treatment of the projectile and analyzed the trade-off for constructing the block encoding with and without amplitude amplification for the potential state preparation. Specifically, this required modifications to the potential state preparation, additional kinetic energy preparation analysis, and new LCU $1$-norm values. We expect that these modifications can serve as the basis for other mixed non-BO simulations in first quantization. For the product formula analysis we introduced a newly optimized eighth-order formula based on the numerical protocol described in Reference~\cite{morales2022greatly} and greatly improved the algorithmic implementation for computing the inverse square root, the most expensive step of the propagator construction.  Here we have improved this step by using QROM function approximation from Reference~\cite{Sanders2020} followed by a single step of Newton-Raphson iteration. In order to analyze the overall effect of these subroutine improvements we derived the total number of product formula steps required for fixed precision and analyzed constant factors by numerically computing the spectral norm of the difference between the product formula and the exact unitary through an adapted power-iteration algorithm. This worst-case bound indicated two-orders of magnitude reduction in the Toffoli complexity for time evolution.

To complete our cost estimates for the stopping power we explored two different projectile kinetic energy estimation strategies. One involved sampling the kinetic energy of the projectile via Monte Carlo mean estimation and the second involved a Heisenberg scaling algorithm, developed in Reference~\cite{kothari2023mean}, with a quadratic improvement over generic sampling. For the Monte Carlo sampling strategy we utilized classical TDDFT to numerically determine an error bound, and required number of samples, for low-Z projectiles. In the KO-algorithm case, we provided the first constant factor analysis of the algorithm's core primitives leveraging Cirq-FT~\cite{khattar_tanuj_2023_8161656}'s resource estimation functionality. While stopping power for ICF targets turns out to only require standard limit Monte Carlo mean estimation there are a number of other settings where stopping power estimation with different accuracy parameters can be useful. In those cases, additional constant factor analysis would be required to analyze the additional classical reduction and state preparation overheads.

This work adds to the body of literature seeking to articulate specific real-world problems of high value where quantum computing might have a large impact, and to quantify the magnitude of advantages offered by fault-tolerant quantum computers. As previous work in this area has shown it is not always straightforward to identify scientific problems amenable to large quantum speedups~\cite{rubin2023fault}. However, recent work suggests that when one is interested in exact electron dynamics -- perhaps the most natural simulation problem for quantum computers -- asymptotic speedups are possible over even computationally efficient mean-field classical strategies~\cite{babbush2023meanfield}. These speedups are even more pronounced at finite temperature. Thus, materials properties in the pre-ignition phase of ICF and other applications in the WDM regime are examples from a particularly rich area for exploration. Adding to this argument is the considerable difficulty in simulating flagship scale problems classically. For WDM there are no efficient, systematically improvable classical methods for first-principles electron dynamics. This work has shown that while quantum dynamics on quantum computers are a promising area, large constant factors for systems of practical interest continue to encourage further investigation into problem representation, observable estimation, classical benchmarking, and classical determination of scaling factors.

\subsection*{Acknowledgments}
The authors thank Bill Huggins, Lucas Kocia, Robin Kothari, Alicia Magann, Jarrod McClean, Tom O'Brien, Shivesh Pathak, Antonio Russo, Stefan Seritan, Rolando Somma, and Andrew Zhao for helpful discussions.
We thank Alexandra Olmstead for providing some of the atomic configurations taken from Ref.~\cite{hentschel2023improving}.
DB worked on this project under a sponsored research agreement with Google Quantum AI. DB is also supported by Australian Research Council Discovery Project DP210101367. AK and ADB were supported by Sandia National Laboratories' Laboratory Directed Research and Development Program (Project No. 222396) and the National Nuclear Security Administration's Advanced Simulation and Computing Program. 
This article has been co-authored by employees of National Technology \& Engineering Solutions of Sandia, LLC under Contract No. DE-NA0003525 with the U.S. Department of Energy (DOE). The authors own all right, title and interest in and to the article and are solely responsible for its contents. The United States Government retains and the publisher, by accepting the article for publication, acknowledges that the United States Government retains a non-exclusive, paid-up, irrevocable, world-wide license to publish or reproduce the published form of this article or allow others to do so, for United States Government purposes. The DOE will provide public access to these results of federally sponsored research in accordance with the DOE Public Access Plan \url{https://www.energy.gov/downloads/doe-public-access-plan}.
Some of the discussions and collaboration for this project occurred while using facilities at the Kavli Institute for Theoretical Physics, supported in part by the NSF under Grant No. NSF PHY-1748958.

\bibliography{main,references}

\clearpage
\begin{center}
\textbf{\large Supplementary Information:\\ Quantum computation of stopping power for inertial fusion target design}
\end{center}

The supplementary information includes the following.
\begin{itemize}
 \item Appendix~\ref{app:accounting_nucleus_pw} accounts for implementation details related to using a larger plane-wave basis set for the projectile nucleus than the target electrons.
 \item Appendix~\ref{app:knockout_algorithm} provides constant factor resource estimates for measuring the projectile kinetic energy with the  algorithm in Ref.~\cite{kothari2023mean}.
 \item Appendix~\ref{app:cost_of_trotter} provides details on the $8^{\rm th}$-order product formula used in this work.
 \item Appendix~\ref{app:precision_stopping} provides numerical justification for the various size of sampling errors for stopping power estimates.
\end{itemize}

\appendix
\section{Accounting for a larger number of plane waves for the nucleus}\label{app:accounting_nucleus_pw}

We allow a different (larger) set of momenta for the projectile than for the electron. The complete Hamiltonian is therefore slightly modified from that in \cite{Su2021} to
\begin{align}
    H &= T_{\rm elec} + T_{\proj} + U_{\rm elec} + U_{\proj} + V_{\rm elec} + V_{\rm elec-proj} \\ 
    T_{\rm elec} &= \sum_{i=1}^{\eta}\sum_{p \in G}\frac{\|k_{p}\|^{2}}{2}|p\rangle\langle p|_{i} \\
    T_\proj &= \sum_{p\in \tilde{G}} \frac{\left \| k_p - k_{\rm proj}\right\|^2}{2 \, M_{\proj}} \ket{p}\!\bra{p}_{\proj} \\
    U_{\rm elec} &= -\frac{4\pi}{\Omega}\sum_{\ell=1}^{L}\sum_{i=1}^{\eta}\sum_{\substack{p,q \in G\\p\neq q}} \left(\zeta_{\ell}\frac{e^{ik_{q - p}\cdot R_{\ell}}}{\|k_{q-p}\|^{2}} \right)|p\rangle \langle q|_{i}\\
    U_\proj  &= \frac{4 \pi}{\Omega} \sum_{\ell=1}^{L}\sum_{\substack{p,q\in \tilde{G} \\ p\neq q}}\bigg(\zeta_\ell\zeta_{\proj}\frac{e^{ik_{q-p}\cdot R_\ell}}{\norm{k_{p-q}}^2}\bigg)\ket{p}\!\bra{q}_{\proj} \\
    V_{\rm elec} &= \frac{2\pi}{\Omega} \sum_{i\neq j}^{\eta} \sum_{p, q \in G} \sum_{\substack{\nu \in G_{0} \\ (p + \nu) \in G \\ (p - \nu) \in G }} \frac{1}{\|k_{\nu}\|^{2}} |p + \nu\rangle \langle p|_{i} |q - \nu\rangle \langle q|_{j} \\
	V_{\rm elec-proj} &= -\frac{4 \pi}{\Omega} \sum_{i=1}^{\eta}\sum_{\substack{p \in G \\ q \in \tilde{G}}} \sum_{\substack{\nu\in G_0\\(p+\nu)\in G \\ (q - \nu) \in \tilde{G}}}\frac{\zeta_\proj}{\left\| k_{\nu}\right\|^2} |p + \nu\rangle\langle p|_{i} |q - \nu\rangle \langle q|_{\proj} 
\end{align}
where the subscript $\proj$ is used to indicate quantities for the projectile nucleus treated quantum mechanically and $\ell$ indexes the nuclei treated within the BO approximation. The set $\tilde{G}$ is now the momenta for the projectile,
$U_\proj$ is used for the potential energy between the single projectile treated quantum mechanically and the other nuclei, and $V_{{\rm elec}-\proj}$ is used for the potential energy between the projectile and the electrons.

There will need to be a different preparation of a $1/\|\nu\|$ state for $U_\proj$ than for the other potential operators,
 because it will need differences over the full range of projectile momentum.
For $V_\proj$ we need to check that $(q-\nu)\in \tilde{G}$, which means that the projectile momentum has not been shifted outside the range of its allowed values.
The sum over $\nu$ here is still over $G_0$, because that includes all allowable shifts of momenta for electrons. Furthermore, we still have the condition that $(p+\nu)\in G$ for the electron momentum to not be shifted outside the allowable range.
The projectile kinetic energy $T_\proj$ differs from that for electrons in that it has division by $M_{\proj}$ for the mass of the projectile (with units chosen such that the electron mass is 1), 
as well as the sum over $\tilde{G}$.
The terms $U_\proj$ and $V_\proj$ have factors for the charge of the projectile, $\zeta_\proj$.

To improve the efficiency, we consider the case where the projectile momentum is centred around some offset.
Then instead of $\left \| k_p\right\|^2$, we would have $\left \| k_p - k_{\rm proj}\right\|^2$.
We need to add $\sum_{w\in\{x,y,z\}}[(k_{\rm proj}^w)^2-2k_p^wk_{\rm proj}^w]$ where the superscript $w$ is indicating the Euler direction.
Since $(k_{\rm proj}^x)^2$ is a classically chosen number, it just gives an undetectable global phase shift which can be ignored, and we just need $-2k_p^wk_{\rm proj}^w$ (the $w$-components of $k_p$ multiplied by a constant).
This means that we are effectively adding an extra term to the Hamiltonian
\begin{equation}\label{eq:kmean_nuclei_appendix}
    T_{\rm mean} = - \sum_{w\in\{x,y,z\}} \sum_{p\in \tilde{G}} \frac{k_p^w k_{\rm proj}^w }{M_{\proj}} \ket{p}\!\bra{p}_\proj \, .
\end{equation}

We will now analyse the complexity for the block encoding of this Hamiltonian.
We will not analyse the interaction picture approach, which is likely to be more costly. 

\subsection{Value of $\lambda$}

First, we will define a new value corresponding to the sum of $1/\|\nu\|^2$ over the wider range as
\begin{equation}
    \lambda_\nu^\proj = \sum_{\nu\in \tilde{G}_0} \frac 1{\norm{\nu}^2}.
\end{equation}
Here $\tilde{G}_0$ is the equivalent of $G_0$, except for differences between elements of $\tilde{G}$ for the projectile momentum.
The contributions to $\lambda$ from $U_{\rm elec}$ and $U_\proj$ are
\begin{align}
    \lambda_{U}^{\rm elec} &= \frac{\eta \lambda_\zeta}{\pi\Omega^{1/3}}\lambda_\nu\, , \nn 
    \lambda_U^\proj &= \frac{\zeta_\proj \lambda_\zeta}{\pi\Omega^{1/3}}\lambda_\nu^\proj\, ,
\end{align}
respectively.
For $U_\proj$, we need $\nu$ summed over $\tilde{G}_0$ for differences of projectile momentum.
The contributions to $\lambda$ from $V$ and $V_\proj$ are
\begin{align}
    \lambda_{V}^{\rm elec} &= \frac {\eta(\eta-1)}{2\pi \Omega^{1/3}}\lambda_\nu\, , \nn 
    \lambda_{V}^\proj &= \frac {\eta\zeta_\proj}{\pi \Omega^{1/3}}\lambda_\nu \, ,
\end{align}
respectively.

The contribution to $\lambda$ from the electron component of the kinetic energy is
\begin{equation}
\lambda_T = \frac{6\eta\pi^2}{\Omega^{2/3}} 2^{2(n_p-1)} .
\end{equation}
As discussed in \cite{Su2021}, the reason why there is the square of $2^{n_p-1}$ rather than $2^{n_p-1}-1$ is because there is a simplification in the state preparation for the registers selecting the bits of the momentum.
The component of $\lambda$ for the projectile, but ignoring the mean is
\begin{equation}
    \lambda_T^{\proj} = \frac{6\pi^2}{M_{\rm proj}\Omega^{2/3}} 2^{2(n_n-1)} .
\end{equation}
The component of $\lambda$ for the product of the offset and the mean will then be obtained from $k_{\max}k^w_{\rm proj}$ for component $w$, where the factor of 2 from squaring and the factor of $1/2$ for kinetic energy cancel.
Because $k=2\pi p/\Omega^{1/3}$, that gives
\begin{equation}
    \frac{2\pi \sum_{w\in\{x,y,z\}} |k^w_{\rm proj}|}{M_{\rm proj}\Omega^{1/3}} 2^{n_n-1} .
\end{equation}
Here we have accounted for the state preparation giving an effective $2^{n_n-1}$ rather than $2^{n_n-1}-1$.
This will also be needed for implementing $k^w_{\rm proj}$ (because it will effectively correspond to all ones classically), so the cost will need to be adjusted by a factor of $2^{n_n-1}/(2^{n_n-1}-1)$.
That gives
\begin{equation}
    \lambda_T^{\rm mean} = \frac{2\pi \sum_{w\in\{x,y,z\}}|k^w_{\rm proj}|}{M_{\rm proj}\Omega^{1/3}} \frac{2^{2(n_n-1)}}{2^{(n_n-1)}-1} .
\end{equation}

Now consider the value of $\lambda$ as given in Eq.\ (119) of \cite{Su2021}, which is
\begin{equation}
    \lambda = \max\left[ \lambda_T+ \lambda_U+\lambda_V, [\lambda_U+\lambda_V/(1-1/\eta)]/p_\nu \right] .
\end{equation}
The reason for this equation is that, in the case where the inequality test $i\ne j$ fails, or the preparation of the $1/\|\nu\|$ state fails, one can simply apply the kinetic energy component of the Hamiltonian.
In the case where that would yield a larger contribution to $T$ than the actual size, that would imply you need to perform an AND with a qubit flagging $T$, and flag a result of 0 as `failure' (removing that contribution to the block encoding).
When considering the effective $\lambda$ values with failures of state preparation, it is divided by the probability of success, so we would have $[\lambda_U+\lambda_V/(1-1/\eta)]/p_\nu$.
For further explanation see Ref.~\cite{Su2021} or Appendix \ref{selstateprep}.
In the case where it would not yield a sufficient contribution to $T$, there would need to be application of $T$ based on an OR with a qubit flagging $T$.
That would imply that $\lambda_T+ \lambda_U+\lambda_V$ is the correct value of $\lambda$ to use.

In our case, the only contribution to the Hamiltonian where we would apply $T_{\elec}$ (or $T_\proj$ or $T_{\rm mean}$) if $i=j$ is if we were otherwise applying $V_{\elec}$, which corresponds to $\lambda_{V}^{\elec}$.
Then for the preparation of the $1/\|\nu\|$ state, we have a distinct preparation for $U_\proj$ (corresponding to $\lambda_U^\proj$) as for the other contributions to the potential energy.
We would therefore have $1/(1-1/\eta)$ for $\lambda_V^{\elec}$ alone, and $1/p_\nu$ for most potential terms, except $1/p^\proj_\nu$ for $U_\proj$.
Therefore the new expression for $\lambda$ is
\begin{equation}\label{eq:lamval}
    \lambda = \max\left[ \lambda_T^{\elec} +\lambda^{\proj}_T+\lambda^{\rm mean}_T+ \lambda_U^{\elec} +\lambda^\proj_U+\lambda_V^{\elec} +\lambda^\proj_V, [\lambda_U^{\elec} +\lambda_V^{\elec}/(1-1/\eta)+\lambda^\proj_V]/p_\nu +\lambda_U^\proj/p_{\nu,\proj} \right] .
\end{equation}
This expression will be discussed in more detail below where we analyse the state preparation.
In the case where amplitude amplification is used for the $1/\|\nu\|$ state preparation, there will be a similar expression with $p_\nu$ and $p_{\nu,\proj}$ replaced with the corresponding probabilities with the amplitude amplification.

\subsection{Preparation cost}
We now need to have a separate superposition over $\nu$ prepared for $U_\proj$ than for all other potential terms, and there will need to be a different preparation over the bits of $T$ for the projectile and electron momenta.
We will also need to adjust the preparation of the registers for selecting between the different terms in the Hamiltonian.

\subsubsection{Preparation of $\nu$ state}
First note that the most difficult part of the preparation is that we need different superpositions over $\nu$ depending on whether we have $U_\proj$ or any other part of the Hamiltonian.
    Referring to Eq.\ (77) of \cite{Su2021}, the first step in the preparation via nested boxes is to prepare a state of the form
    \begin{equation}
        \frac 1{\sqrt{2^{n_p+2}}} \sum_{\mu=2}^{n_p+1} \sqrt{2^\mu} \ket{\mu},
    \end{equation}
    where $\ket{\mu}$ is encoded in unary.
    In our case we will need the equivalent state except with $n_p$ replaced with $n_n$ for the case of $U_\proj$.
    Because the state is prepared by a sequence of controlled Hadamards, one can control between preparing the two states by making $n_n-n_p$ of the Hadamards also controlled by the qubit selecting $U_\proj$.
    This just increases the cost of the controlled Hadamards by 1 each for an extra cost of $n_n-n_p$ Toffolis.
    
    The useful feature of this approach is that \emph{no} further amendment to the preparation scheme is needed to make it controlled.
    The rest of the state preparation for $\nu$ can proceed exactly as before, with the only extra Toffoli cost being $n_n-n_p$ at the beginning for preparing the nested boxes state.
    
    To explain the controlled preparation scheme in more detail, the unary encoding is of the form
     \begin{equation}
        \frac 1{\sqrt{2^{n_p+2}}} \sum_{\mu=2}^{n_p+1} \sqrt{2^\mu} \ket{\mu} =
        \frac 1{\sqrt{2^{n_p+2}}} \sum_{\mu=2}^{n_p+1} \sqrt{2^\mu} \ket{0\cdots 0 \underbrace{1 \cdots 1}_{\mu}} \, .
    \end{equation}
    The unary basis state corresponding to $\mu = n_p+1$ corresponds to $\ket{1\cdots 1}$.
    The start of the state preparation is to perform a Hadamard on the first qubit, then use that to control a Hadamard on the second qubit, and so forth.
    At the end we would perform a controlled Hadamard on the second-last qubit.
    This would give an equal superposition between $\mu=2$ and $\mu=1$, but because we do not allow $\mu<2$, the case $\mu=1$ would be flagged as a failure.
    The final qubit depicted here can be omitted, because it would always be 1 in this encoding.
    There will be $n_p$ qubits, and $n_p-1$ controlled Hadamards.
    
    In our case here, we would want to either prepare this state or
     \begin{equation}
        \frac 1{\sqrt{2^{n_n+2}}} \sum_{\mu=2}^{n_n+1} \sqrt{2^\mu} \ket{0\cdots 0 \underbrace{1 \cdots 1}_{\mu}} \, ,
    \end{equation}
    where $n_p$ has been replaced with $n_n$.
    Because these states need to be represented on the same qubits, we would have $n_n-n_p$ leading zeros when preparing the first state.
    Therefore, for the first Hadamard on the first qubit, it would need to be controlled on the qubit selecting between the two states.
    Then for the next qubit, provided $n_n-n_p>1$, we would perform a doubly-controlled Hadamard.
    That is, the Hadamard on the second qubit would be controlled by the first qubit and the qubit selecting between the two states.
    This will be true for all following qubits that need to be zero for the $n_p$ state.
    Making the Hadamard on the first of these qubits controlled, and the remaining Hadamards doubly controlled, gives an extra Toffoli complexity of $n_n-n_p$.
    
    For the first qubit that is non-zero for the $n_p$ state, we would need to perform a Hadmard for that state, or a controlled Hadmard for the $n_n$ state.
    This selection does not require any further non-Clifford gates.
    One can simply use the qubit selecting the $n_p$ state as the control for a CNOT on the preceding qubit.
    That ensures it is 1 for the $n_p$ state.
    Then perform the controlled Hadamard as before.
    For the $n_n$ case this is just part of the sequence of controlled Hadamards, but for $n_p$ it is ensuring the Hadamard is performed on the qubit.
    Then just perform another CNOT to erase the preceding qubit.
    Then the sequence of controlled Hadamards can proceed in the same way as when not preparing this state in a controlled way.
    By this procedure one can control between preparing the state with $n_p$ or $n_n$ with an extra Toffoli cost of only $n_n-n_p$.
    
\subsubsection{Preparation of momentum control qubits}
Next consider how to prepare superpositions over control qubits for the bits of the momentum.
    This preparation will need to be controlled by a qubit selecting between the electron and projectile momentum.
    The preparation is described in Eqs.\ (67) to (69) of \cite{Su2021}, and again it proceeds by a sequence of Hadamard gates, except this time it needs to be done twice for two states (giving $r$ and $s$).
    We can make this controlled in exactly the same way as for the preparation of $\nu$.
    The only difference is that this time there are two states, so the extra Toffoli cost is $2(n_n-n_p)$.
    
    We will also need a preparation for a state selecting between the components of $k_{\rm proj}$ for the product $k_p^w k_{\rm proj}^w$.
    In practice, the direction of $k_{\rm proj}$ does not need to be taken to be very precise, and we can just consider a rounded direction.
    We will therefore just use 8 bits for selecting between the components.
    The exact value chosen has very little effect on the overall cost.
    The method is to use an 8-qubit equal superposition state (prepared with Hadamards).
    There are then two inequality tests to prepare the qubits for selecting between $x,y,z$, which a total cost of 16 Toffolis.
    These can be inverted with Cliffords for the inverse state preparation provided the temporary qubits are retained.
    
    We will also need to use the qubit selecting the product of the mean momentum and offset to control a swap of these qubits and those that are used for selecting $x,y,z$ for the square of the momentum.
    That will cost another 4 Toffolis, including 2 for the controlled swap and another 2 for the inversion.
    
\subsubsection{Preparation of state selecting term in Hamiltonian}
\label{selstateprep}

In practice we are applying the kinetic component of the Hamiltonian in the case of failure of state preparation for the potential terms.
In particular, there are two scenarios, corresponding to whether the first or second expression in Eq.\ \eqref{eq:lamval} gives the maximum.
When the first is larger, this implies that only applying the kinetic term in the case of state preparation failure will not give sufficient weight on that term.
You need to apply the kinetic term if there is failure OR a qubit flagging the kinetic term is in the $\ket{1}$ state.
That can be computed with a Toffoli.
In the case where the second expression in Eq.\ \eqref{eq:lamval} is larger, that means that applying the kinetic term in the case of preparation failure would give too large a weight on the kinetic term.
Then one needs to apply the kinetic term if there is failure AND the qubit flagging the kinetic term is in the $\ket{1}$ state.
That logical AND is again something that can be computed with a single Toffoli.

But, unlike in \cite{Su2021} there are three kinetic components to account for.
This means that, in the case of failure of the state preparation we also need a register to select between the three kinetic components.
To achieve this, we will prepare a state of the form
\begin{equation}
    \left( \sqrt{\alpha_{UV}} \ket{0} + \sqrt{\alpha_T} \ket{1} \right) \left( \sqrt{\mu_T^{\elec}} \ket{0} + \sqrt{\mu^\proj_T} \ket{1} + \sqrt{\mu^{\rm mean}_T} \ket{2} \right) \left( \sqrt{\mu_U^{\elec}} \ket{0} + \sqrt{\mu^\proj_U} \ket{1} + \sqrt{\mu_V^{\elec}} \ket{2} + \sqrt{\mu^\proj_V} \ket{3} \right)\, ,
\end{equation}
where the first qubit is used to select the kinetic component, the second register is to select between the different kinetic energy components, and the third register is used to select between the potential energy components.

Now, in the case where the first expression in Eq.\ \eqref{eq:lamval} gives the maximum, we would perform an OR between the result of state preparation and the first qubit, and use the second register to select between the components of $T$.
To describe this state preparation in a simplified way, we will describe it as a rotated qubit flagging success.
It will, of course, be entangled with the prepared state, but we are ignoring that for the simplicity of the explanation here.
The state can then be written as
\begin{align}
    &\left( \sqrt{\alpha_{UV}} \ket{0} + \sqrt{\alpha_T} \ket{1} \right) \left( \sqrt{\mu_T^{\elec}} \ket{0} + \sqrt{\mu^\proj_T} \ket{1} + \sqrt{\mu^{\rm mean}_T} \ket{2} \right) \left[ \left(\sqrt{\mu_U^{\elec}} \ket{0} + \sqrt{\mu^\proj_V} \ket{3}\right) \left( \sqrt{p_\nu}\ket{0}+\sqrt{1-p_\nu}\ket{1}\right) \right. \nonumber \\
    & \quad
\left. + \sqrt{\mu^\proj_U} \ket{1}\left( \sqrt{p_{\nu,\proj}}\ket{0}+\sqrt{1-p_{\nu,\proj}}\ket{1}\right) + \sqrt{\mu_V^{\elec}} \ket{2} \left( \sqrt{(1-1/\eta)p_\nu}\ket{0}+\sqrt{1-(1-1/\eta)p_\nu}\ket{1}\right) \right] \, .
\end{align}
This corresponds to a probability of $p_\nu$ for success with $U$ or $V_\proj$, since we only need to prepare the $1/\|\nu\|$ state with $n_p$ qubits.
Then there is $p_{\nu,\proj}$ for $U_\proj$ since there is preparation with $n_n$ qubits for the projectile.
Lastly, for $V$ there is $(1-1/\eta)p_\nu$ since we need preparation of the $1/\|\nu\|$ state and $i\ne j$ in preparing the equal superposition state.

To take account of the case where there is amplitude amplification performed for the state preparation for $\nu$, there will be separate boosted probabilities $p_\nu^{\rm amp}$ and $p_{\nu,\proj}^{\rm amp}$ for the electron and projectile parts.
This is because the state preparation and amplitude amplification is performed entirely controlled by the register selecting between electron and projectile components.
(A different expression would be obtained if there were amplitude amplification involving the selection between components as well.)
We give the reasoning below using the expressions for the un-amplified probabilities, but exactly the same reasoning applies with the amplified probabilities.

The overall squared amplitude for $\ket{0}$ on the ancilla flag qubit is then
\begin{equation}
    \mu_{UV}:=p_\nu (\mu_U^{\elec} +(1-1/\eta) \mu_V^{\elec}+ \mu^\proj_V)+ p_{\nu,\proj} \mu^\proj_U \, .
\end{equation}
We would \emph{only} apply a potential component of the Hamiltonian if we have $\ket{0}$ on this qubit and $\ket{0}$ on the first qubit, which has a squared amplitude $\alpha_{UV}$.
Therefore the squared amplitude for performing the kinetic component of the Hamiltonian at all is $1-\alpha_{UV}\mu_{UV}$.
The squared amplitudes for applying the kinetic components will correspond to this factor times the squared amplitudes in the second register.
So, for example, the squared amplitude for $T_{\elec}$ is $(1- \alpha_{UV} \mu_{UV}) \mu_T^{\elec}$.

In the overall block encoding, the block encoding for the Hamiltonian gives $H/\lambda$, which is how $\lambda$ is defined.
Here we would have a squared amplitude for $T_{\elec}$, then block encode $T_{\elec}$ with a factor of $1/\lambda_T^{\elec}$.
This means that the factor of $1/\lambda$ in the overall block encoding needs to be the same as $(1- \alpha_{UV} \mu_{UV}) \mu_T^{\elec}/\lambda_T^{\elec}$.
Solving for $\lambda_T^{\elec}$ then gives
\begin{equation}
    \lambda_T^{\elec} = \lambda \left(1- \alpha_{UV} \mu_{UV}\right) \mu_T^{\elec} \, .
\end{equation}
In exactly the same way, for the other kinetic components we obtain
\begin{equation}
    \lambda^{\proj}_T  = \lambda \left(1- \alpha_{UV} \mu_{UV}\right) \mu^{\proj}_T \, , \qquad
    \lambda^{\rm mean}_T  = \lambda \left(1- \alpha_{UV}\mu_{UV}\right) \mu^{\rm mean}_T \, .
\end{equation}
Then, for the potential components, we need to multiply the squared amplitude for that component in the third register by the squared amplitudes for $\ket{0}$ on the first register and flag qubit.
For example, for $U_{\elec}$ we have
\begin{equation}
\lambda_U^{\elec} = \lambda\, \alpha_{UV} p_\nu \mu_U^{\elec} \, ,
\end{equation}
where the factor of $\lambda$ arises from exactly the same reasoning as for the kinetic components.
Similarly, for the other potential components we have
\begin{equation}
    \lambda^\proj_U = \lambda\, \alpha_{UV} p_{\nu,\proj} \mu^\proj_U \, , \qquad
    \lambda_V^{\elec} = \lambda\, \alpha_{UV} p_\nu (1-1/\eta) \mu_V^{\elec} \, , \qquad
    \lambda^\proj_V = \lambda\, \alpha_{UV} p_\nu \mu^\proj_V \, .
\end{equation}
Next, note that when we sum $\lambda_T^{\elec},\lambda_U^{\elec},\lambda_V^{\elec},\lambda_T^\proj,\lambda_U^\proj,\lambda_V^\proj,\lambda_T^{\rm mean}$, we just get $\lambda$.
That is what is expected, because there is no case here where no component of the Hamiltonian is implemented.
This means that the individual $\lambda$ values should correspond to sums of weights in linear combinations of unitaries for the components of the Hamiltonian, with their sum corresponding to the sum of weights for the complete Hamiltonian.
This is also consistent with the first expression in Eq.\ \eqref{eq:lamval}.

Let us prepare the registers using inequality tests using numbers of qubits $n_{VT}$, $n_T$, and $n_{UV}$, for the respective registers.
There will then need to be six inequality tests for the preparation.
First, the possible error in $\alpha_T$ or $\alpha_{UV}$ will be $\lambda/2^{n_{VT}+1}$.
The total contribution to the error in $\lambda_T^{\elec},\lambda^{\proj}_T,\lambda^{\rm mean}_T$ will come from the sum of that in the expressions for those three quantities.
The contribution to the error should be no more than
\begin{equation}
    \frac{\lambda}{2^{n_{VT}+1}}\mu_{UV} \, .
\end{equation}
We get this exact same expression if we sum the possible contributions to the error from $\lambda_U^{\elec},\lambda^\proj_U,\lambda_V,\lambda^\proj_V$.
We therefore find that the contribution to the error from this source (preparation of the first qubit) is no larger than 
\begin{equation}
    \frac{\lambda}{2^{n_{VT}}}\mu_{UV} \le \frac{\lambda}{2^{n_{VT}}} \, .
\end{equation}

Next, consider the contribution to the error from the imprecision in the inequality tests for the second register.
The error in $\mu_T^{\elec},\mu^{\proj}_T,\mu^{\rm mean}_T$ can be $1/2^{n_{T}+1}$ for two, and $1/2^{n_{T}}$ for the third, for a total of $2/2^{n_{T}}$.
This error is multiplied by $\lambda (1- \alpha_{UV} \mu_{UV})$, to give
\begin{equation}
     \frac{2\lambda}{2^{n_{T}}}\left(1- \alpha_{UV}\mu_{UV}\right) \, .   
\end{equation}
Then, the contributions to the error from the inequality tests for the third register are $1/2^{n_{UV}+1}$ for two, and $1/2^{n_{UV}}$ for two.
The total contribution to the error will then be smaller than
\begin{equation}
     \frac{\lambda}{2^{n_{UV}}}\alpha_{UV}\mu_{UV} \, .   
\end{equation}
Now, if we take $n_{UV}=n_{T}-1$, then the total of the error from the two sources (preparation of the second and third registers) is upper bounded by
\begin{equation}
    \frac{2\lambda}{2^{n_{T}}} .
\end{equation}
If we take $n_{VT}=n_{T}$, then the total error from the three sources is upper bounded as
\begin{equation}
    \frac{3\lambda}{2^{n_{T}}} .
\end{equation}

The other case is that when the second expression in Eq.\ \eqref{eq:lamval} gives the maximum.
That corresponds to the failure cases of the state preparation giving a weighting that is too large for $T$, so an AND is performed with the qubit flagging $T$ to reduce the weight to the correct value.
In that case, the same three-register state is used, and the result of the state preparation success flag state is the same.
But, we apply kinetic components only if there is a failure of the state preparation (which happens with squared amplitude $1- \mu_{UV}$, and there is $\ket{1}$ on the first qubit, with squared amplitude $\alpha_T$.
By exactly the same reasoning as before, we obtain the individual $\lambda$-values as the squared amplitudes multiplied by $\lambda$.
That gives the kinetic $\lambda$-values as
\begin{equation}
     \lambda_T^{\elec} = \lambda\, \alpha_T  \left(1- \mu_{UV}\right) \mu_T^{\elec} \, , \qquad
    \lambda^{\proj}_T  = \lambda\, \alpha_T  \left(1- \mu_{UV}\right) \mu^{\proj}_T \, , \qquad
    \lambda^{\rm mean}_T  = \lambda\, \alpha_T \left(1- \mu_{UV}\right) \mu^{\rm mean}_T \, .
\end{equation}
For the potential energy components, we apply them \emph{whenever} there is success of the state preparation, regardless of the state of the first qubit.
This means that the $\lambda$-values are
\begin{equation}
    \lambda_U^{\elec} = \lambda\,  p_\nu \mu_U^{\elec} \, , \qquad
    \lambda^\proj_U = \lambda\,  p_{\nu,\proj} \mu^\proj_U \, , \qquad
    \lambda_V^{\elec} = \lambda\,  p_\nu (1-1/\eta) \mu_V^{\elec} \, , \qquad
    \lambda^\proj_V = \lambda\,  p_\nu \mu^\proj_V \, .
\end{equation}
It is then easy to check that
\begin{equation}
    [\lambda_U^{\elec}+\lambda_V^{\elec}/(1-1/\eta)+\lambda^\proj_V]/p_\nu +\lambda_U^\proj/p_{\nu,\proj} = \lambda \, ,
\end{equation}
so the second expression in Eq.\ \eqref{eq:lamval} gives $\lambda$ as expected.
Note that this is the expression without amplitude amplification for $\nu$.
When amplitude amplification is used the probabilities $p_\nu$ and $p_{\nu,\proj}$ are replaced with the amplified probabilities $p_\nu^{\rm amp}$ and $p_{\nu,\proj}^{\rm amp}$.

The error in $\alpha_T$ will only affect $\lambda_T^{\elec},\lambda^{\proj}_T,\lambda^{\rm mean}_T$, because the implementation of the potential energy components is independent of the first qubit.
Given that the error in $\alpha_T$ is upper bounded as $1/2^{n_{VT}+1}$, the contribution to the error in the three kinetic terms is upper bounded as
\begin{equation}
    \frac{\lambda}{2^{n_{VT}+1}}\left(1- \mu_{UV}\right) \, .
\end{equation}
Next consider the contribution to the error from the preparation of the second register.
The total error in $\mu_T^{\elec},\mu^{\proj}_T,\mu^{\rm mean}_T$ can be upper bounded as $2/2^{n_T}$, to give an upper bound on the contribution to the error in $\lambda_T^{\elec},\lambda^{\proj}_T,\lambda^{\rm mean}_T$ as
\begin{equation}
     \frac{2\lambda}{2^{n_{T}}}\alpha_T\left(1- \mu_{UV}\right) \, .
\end{equation}
The contributions to the error from the inequality tests for the third register will then be
\begin{equation}
     \frac{\lambda}{2^{n_{UV}}}[p_\nu (\mu_U^{\elec} +(1-1/\eta) \mu_V^{\elec}+ \mu^\proj_V)+ p_{\nu,\proj} \mu^\proj_U] = \frac{\lambda}{2^{n_{UV}}}\mu_{UV}  \, . 
\end{equation}
The only difference from the first case (using an AND) is that we have removed the factor of $\alpha_{UV}$.
That is because the potential energy component does not depend on the first qubit.
Provided we take $n_{UV}=n_{VT}+1$, adding this to the contribution to the error from $\alpha_T$ gives
\begin{equation}
    \frac{\lambda}{2^{n_{UV}}} .
\end{equation}
If we take $n_{UV}=n_{T}$, then the total error from the three sources (preparation on the three registers) is again upper bounded as
\begin{equation}
    \frac{3\lambda}{2^{n_{T}}} .
\end{equation}

When performing the inequality tests for the state preparation on each register, we naturally obtain a result encoded in unary.
This is we will have one alternative where none of the inequalities are satisfied, giving $000$, another with one satisfied to give $001$, another with two satisfied giving $011$, and so forth.
The encoding of the registers in unary is convenient because we have separate qubits flagging each of the component of the Hamiltonian (after converting to one-hot unary).

In our implementation we will need a qubit selecting between electron and projectile components (so, for example, between $V_{\elec}$ and $V_\proj$).
It is trivial to prepare such qubits separately for the kinetic and potential registers without Toffolis.
But, we will also need to select between these qubits based on a qubit we prepare (discussed further below) selecting between the kinetic and potential components.
That can be performed via a single controlled swap, with another controlled swap in the inverse preparation for a total of two Toffolis.

\subsection{Selection cost}

For the storage of the state, a larger number of qubits $3n_n$ are used for the projectile, but the electron momenta are stored in the same way as in \cite{Su2021}.
The projectile state is stored in a given location, with the antisymmetrization only for the electron registers.
Then, when we are performing the controlled swap of the momentum register to an ancilla, there will be two, one controlled by $i$ and the other by $j$.
For the second we only swap electron momenta into the ancilla, so the procedure is identical to that in \cite{Su2021}.
For the first, we also control on a qubit for selecting the projectile momentum, and the ancilla will include extra qubits to allow for storage of the projectile momentum.
The controlled swaps can just be performed as before, except in the case that the qubit flagging the momentum component is set we swap the $n_n$ qubits for the projectile momentum.
We are performing operations on the ancilla register in superposition (over the registers selecting the electron register or projectile register).
The qubits in this ancilla will all be zeroed, so when we swap in an electron momentum the extra qubits that would be used in the case of the projectile momentum are all zero.
This means it is possible to perform most operations on the momentum register in common between projectile and electron momenta.

Next we consider the cost.
In our implementation, we use $i,j$ to index electron registers, with an extra qubit to select between electron and projectile components.
We will perform the controlled swaps of the projectile momentum into only \emph{one} of the temporary registers.
This means the cost of the controlled swaps is $2(\eta-2)$ for the unary iteration for the electron registers, then $2[(\eta+1)-2]$ for iteration over the electron and projectile registers.
That gives a total of $4\eta-6$.
Then cost of the controlled swaps themselves is
\begin{equation}
    12\eta n_p + 6 n_n .
\end{equation}
This is just an extra cost of $6 n_n$ for the projectile momentum, for two controlled swaps on $3n_n$ qubits.
That gives a cost
\begin{equation}
    12\eta n_p + 6 n_n +4\eta-6 .
\end{equation}

The other main change to the select operations is that the operations need to be on $n_n$ qubits rather than $n_p$.
This will impact the selection cost of the components of the Hamiltonian in different ways, detailed in the next subsection.
There also needs to be selection between the square of the momentum and the product of the momentum with the momentum offset for the kinetic energy.

To see how to modify the procedure to perform this selection, recall how the kinetic energy is computed as in Eq.~(65) of \cite{Su2021}.
There the kinetic energy is written as
\begin{equation}
    T = \frac{2\pi^2}{\Omega^{2/3}} \sum_{j=1}^\eta \sum_{p\in G} \sum_{w\in\{x,y,z\}} \sum_{r=0}^{n_p-2} \sum_{s=0}^{n_p-2} 2^{r+s} p_{w,r} p_{w,s} \ket{p}\!\bra{p}_j \, .
\end{equation}
Here the two bits $p_{w,r} p_{w,s}$ correspond to products of bits in a component of $p$ in order to obtain the square.
The method is very simply modified to obtain the product of the form $k_p^xk_{\rm mean}^x$, simply by selecting $w$ as $x$, and replacing bit $p_{x,s}$ with the corresponding bit for $k_{\rm mean}^x$.
The way we encode $k_{\rm mean}^x$ is that it is represented by all bits equal to 1, with its actual value (the multiplying factor times that integer) being governed by the state preparation.

In particular, we need to modify step 4 in the list of steps in the left column of page 17 of \cite{Su2021}.
This step involves performing a Toffoli controlled by the qubits storing $p_{w,r}$ and $p_{w,s}$.
Here we would make the NOT controlled on $p_{w,r} \wedge (p_{w,s} \vee b)$, for $b$ the flag for the $T_{\rm mean}$ component of the Hamiltonian.

There is also the selection of the component of the momentum depending on whether we want the square or product with the mean.
As explained above, there is a cost of 4 Toffolis.
The register to select the $x,y,z$ component can be given in binary or unary, since we can convert between binary and unary with Cliffords for three.
We can therefore assume that the $w$ registers for the square and product are given using 2 qubits each (for binary).
The qubit selecting $T_{\rm mean}$ can be used to control a swap between these two registers, with a cost of 2 Toffolis, then there are another 2 Toffolis to invert the swap for the inverse state preparation.

There is no extra non-Clifford cost for the minus sign in $T_{\rm mean}$, because it is just a (Clifford) controlled phase gate.
We may also perform controlled phase gates to apply signs of the components of $k_{\rm mean}$ with no extra Toffoli cost.
The condition that $(p+\nu)\in G$ versus $(q-\nu)\in \tilde{G}$ needs no modification.
The operation $p+\nu$ is for an electron momentum, so $(p+\nu)\notin G$ the extra $n_n-n_p$ qubits will not be all zero.
Then those qubits are not swapped back into the momentum register in the controlled swap, so nonzero qubits are remaining to flag a `fail' and remove that part in the block encoding.
Similarly, if $(q-\nu)\notin \tilde{G}$ for the projectile momentum, then there will be an extra ancilla qubit flagging `fail' resulting from the subtraction.
There will be no need to treat these registers differently apart from the controlled swaps addressed above.

\section{Constant factor estimates for projectile kinetic energy measurement using the Knockout algorithm}
\label{app:knockout_algorithm}
For a fixed standard error $\epsilon$, Monte Carlo sampling provides the optimal bound on the number of samples needed to estimate the expected value of an observable.  This \textit{standard} sampling limit states the number of samples to estimate the observable to $\epsilon$ precision goes as $\mathcal{O}(\sigma^{2}/\epsilon^{2})$ where $\sigma^{2}$ is the variance of the observable.  Recently, a quantum algorithm was developed by Kothari and O'Donnell (KO)~\cite{kothari2023mean} that allows one to estimate the expected value of an observable to $\epsilon$ precision with the number of samples going as $\mathcal{O}(\sigma^{2}/\epsilon)$; a quadratic improvement over the standard limit.  The main protocol in this algorithm is the use of phase estimation on a unitary that is the composition of a reflection around the prepared state--called the synthesizer--and a phasing operation that phases basis states according the value the random variable takes on those basis states. This protocol allows one to solve a decision problem that identifies if the expected value, encoded as the eigenvalue of the unitary, is to the left or right of a gapped range, thereby knocking out part of the possible range (Theorem 1.3 in Ref.~\cite{kothari2023mean}).  Using additional classical reductions allows one to boost this decision problem to the mean estimation problem (Theorem 1.1 in Ref.~\cite{kothari2023mean}).  In this section we describe a rough estimate of the constant factors associated with performing the KO algorithm to estimate the kinetic energy of the projectile. We do not detail all classical reductions necessary for the task but instead focus on the primary decision problem to obtain an estimate of when $\epsilon$ is small enough such that the KO algorithm has a computational advantage over Monte Carlo sampling. The unitary that needs to be phase estimated in the KO algorithm is the composition of a reflection analogous to the Grover diffusion operator $\mathrm{REFL}$ and a phase oracle $\mathrm{ROT}_{y}$
\begin{align}
U = \mathrm{REFL} \cdot \mathrm{ROT}_{y}.
\end{align}
Given a circuit that prepares the desired probability distribution
\begin{align}
P|0\rangle = \sum_{\ell}\sqrt{p(\ell)}|\ell\rangle
\end{align}
the reflection is defined as
\begin{align}
\mathrm{REFL} = P\left(2 |0 \rangle \langle 0 | - \mathbb{I}\right)P^{\dagger}
\end{align}
and the $\mathrm{ROT}_{y}$ is
\begin{align}
\mathrm{ROT}_{y}|\ell\rangle = e^{i\alpha_{\ell}}|\ell \rangle
\end{align}
where $\alpha_{\ell} = -2\mathrm{arctan}(y_{\ell})$ and $y_{\ell}$ is the value of the random variable for the event indexed by $\ell$. In order to perform phase estimation on the KO algorithm $U$ we need controlled forms of $\mathrm{REFL}$ and $\mathrm{ROT}_{y}$ along with controlled forms of their inverses. For a non-abridged version of the algorithm and details surrounding allowed ancilla registers see Section 3 of Ref.~\cite{kothari2023mean}. 

For the synthesizer ($P$) in $\mathrm{REFL}$ we use the time-evolution operator. Therefore, building the reflection operator requires two calls to the state previously described state preparation circuit. For the $\mathrm{ROT}_{y}$ operator we first encode the kinetic energy of the projectile into an ancilla register through a series of multiplications and additions on the ancilla and the projectile register.  Second, we calculate the arctan on this register which is linear complexity in the ancilla register size.  Finally, controlled phase gates are used to accomplish the correct action defined by the $\mathrm{ROT}_{y}$ unitary.  The remaining task to get order of magnitude estimates of the quantum resources required for the KO algorithm is to derive a circuit for writing the random variable value to the ancilla register used to compute $\alpha_{\ell}$.

Recall that the kinetic energy operator on the projectile
\begin{align}
T = \sum_{p \in \tilde{G}}\frac{\|k_{p} - k_{\mathrm{proj}}\|^{2}}{2 M_{\mathrm{proj}}}|p\rangle\langle p|
\end{align}
where $\|k_{p} - k_{\mathrm{proj}}\|^{2} = \sum_{w\in\{x,y,z\}}\left((k_{p}^{w})^{2} + (k_{\rm mean}^w)^2-2k_p^wk_{\rm proj}^w\right)$ where $\sum_{w \in \{x,y,z\}} (k_{\rm proj}^{w})^{2}$ is a constant term which we will add with an addition circuit. We can rewrite the coefficients as
\begin{align}
\frac{\|k_{p} - k_{\mathrm{proj}}\|^{2}}{2 M_{\rm proj}} = \frac{1}{2M_{\rm proj}}\left(\frac{2\pi}{\Omega^{1/3}}\right)^{2} \left( \boldsymbol{p}^{T}\boldsymbol{p} - 2\boldsymbol{p}^{T}\boldsymbol{p}_{\rm proj} + \boldsymbol{p}_{\rm proj}^{T} \boldsymbol{p}_{\rm proj}    \right)
\end{align}
where $\boldsymbol{p} = (p_{x}, p_{y}, p_{z})$.  $n_{\rm mean}$ is the number of bits needed to represent the central momentum of the projectile where $n_{\rm mean} > n_{\rm proj}$. If we ignore the constant involving the mass and the volume element we are left with a series of integer products and summations. The Toffoli cost of each of the three terms can be derived from protocols described in the Appendix of Ref.~\cite{Su2021} and are as follows:
\begin{enumerate}
    \item The $\boldsymbol{p}^{T}\boldsymbol{p}$ term involves three $n_{\rm proj}$ registers as thus has $3 n_{\rm proj}^{2} - n_{\rm proj} - 1$ Toffoli complexity.
    \item The $\boldsymbol{p}^{T}\boldsymbol{p}_{\rm proj}$ involves the sum-product of three integer pairs of sizes $(n_{\rm proj}, n_{\rm mean})$.  The products each take $2 n_{\rm mean} n_{\rm proj} - n_{\rm mean}$ and the three sums involved cost $3 n_{\rm mean}^{2} - n_{\rm mean} - 1$. 
    \item The final addition requires $\boldsymbol{p}_{\rm mean}^{T} \boldsymbol{p}_{\rm proj}$ stored in $2 n_{\rm mean} - 1$ bits to be added to the results from steps 1 and 2. if we pad out the results from 1 and 2 up to $2n_{\rm mean} - 1 = n_{f}$ then we need $3n_{f}^{2} - n_{f} - 1$ Toffolis for this operation.
    \item We leave off the constant to be included in multiplying the variance for the mean-estimation algorithm. In the last step we must subtract an estimate of $\mu_{0} = \langle \boldsymbol{p}^{T}\boldsymbol{p} - 2\boldsymbol{p}^{T}\boldsymbol{p}_{\rm proj} + \boldsymbol{p}_{\rm proj}^{T} \boldsymbol{p}_{\rm proj} \rangle$ which we know from classical data. This additional step can be included with the previous step by modifying the subtracted value of $\boldsymbol{p}_{\rm proj}^{T} \boldsymbol{p}_{\rm proj}$ to $\boldsymbol{p}_{\rm proj}^{T} \boldsymbol{p}_{\rm proj} - \mu_{0}$.
\end{enumerate}
Thus the total Toffoli complexity of performing the integer encoding step is 
\begin{align}
C_{\rm encoding} = 3 n_{\rm proj}^{2} - n_{\rm proj} - 1 + 3(2 n_{\rm mean} n_{\rm proj} - n_{\rm mean}) + 3 n_{\rm mean}^{2} - n_{\rm mean} - 1 + 3n_{f}^{2} - n_{f} - 1
\end{align}
These costs combined with the costs associated with two calls to the synthesizer, reflection, and computing the arctan are combined to produce a total Toffoli complexity.  

\section{Bespoke $8^{\rm th}$-order product formula}
\label{app:cost_of_trotter}
The $8^{\rm th}$-order formula we use was numerically determined by solving equations as described in Reference~\cite{Yoshida1990}.
We solved for over 100,000 product formulae, and selected the one with the smallest constant factor via testing with random Hamiltonians of matrix dimension $6 \times 6$.
Further refinement was performed by minimising the Taylor expansion up to $9^{\rm th}$-order (so including the error term) and then solving for the $8^{\rm th}$-order formula.
The method for performing the Taylor expansion is described in Reference~\cite{morales2022greatly}.
The symmetric product formula has the form
\begin{align}
S_{\rm prod} = S_2(w_{10} t) S_2(w_9 t) .... S_2(w_2 t) S_2(w_1 t) S_2(w_0 t) S_2(w_1 t) S_2(w_2
t) .... S_2(w_9 t) S_2(w_{10} t)   
\end{align}
with $w_{0} = 1 - 2\sum_{i=1}^{10}w_{i}$ and $S_2(t) = e^{-itH_{0}/2}e^{-i t H_{1}} e^{-i t H_{0}/2}$ for two non-commuting Hamiltonians $H_{0}$ and $H_{1}$. The following formula was numerically determined
\begin{align}
w=[& 5.935806040085031 \times 10^{-1}, \nonumber \\
-&4.691601234700394 \times 10^{-1}, \nonumber \\
&2.743566425898439 \times 10^{-1},  \nonumber \\
&1.719387948465702 \times 10^{-1}, \nonumber \\
&2.343987448254160 \times 10^{-1}, \nonumber \\
-&4.861642448032533 \times 10^{-1}, \nonumber \\
&4.961736738811380 \times 10^{-1}, \nonumber \\
-&3.266021894843879 \times 10^{-1}, \nonumber \\
&2.327167934936900 \times 10^{-1}, \nonumber \\
&9.824955741471075 \times 10^{-2}]  \nonumber 
\end{align}
and is the bespoke $8^{\rm th}$-order formula we use in this work.

\section{Precision requirements for stopping power \label{app:precision_stopping}}
\begin{figure}[h!]
    \centering
    \includegraphics[scale=0.5]{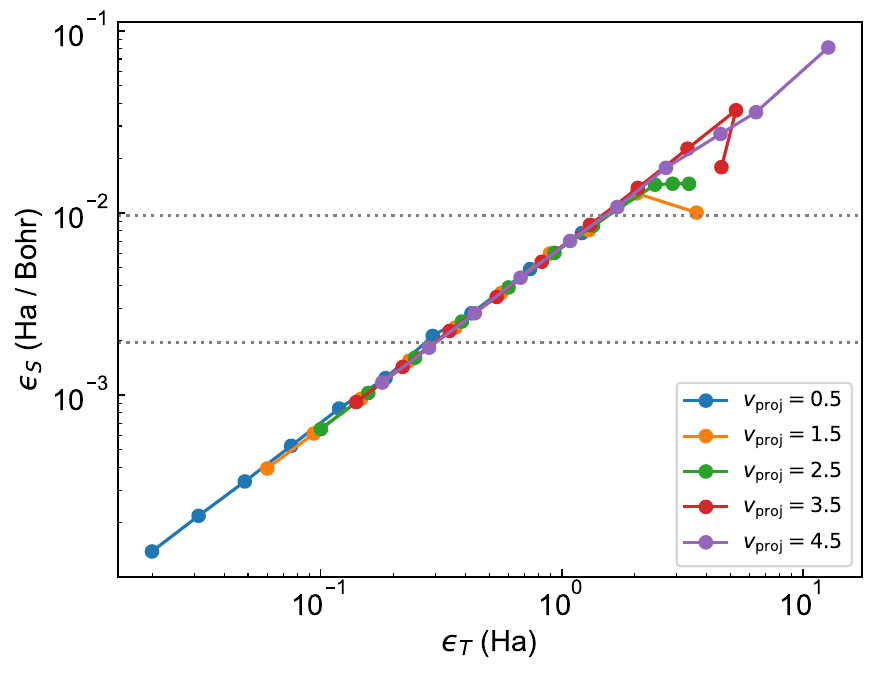}
    \includegraphics[scale=0.5]{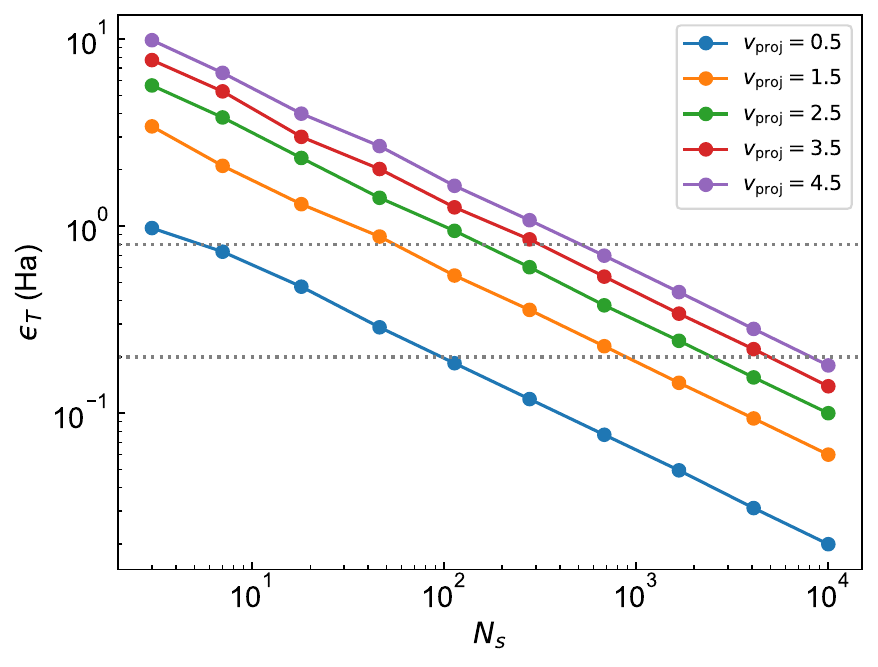}
    \caption{(left)  Dependence of precision in the stopping power $\epsilon_S$ estimate on the precision in the individual kinetic energy data points $\epsilon_T$ for different values of the projectiles initial velocity ($v_\proj$). We extracted the stopping power using 10 equally spaced points and chose $\sigma_k = 4$. Dashed lines represent a desired target precision of $0.1$ and 0.05 eV/\AA. (right) Mean precision in the kinetic energy as a function of the number of samples $N_s$. Dashed lines correspond to the values of $\epsilon_T$ which yield the desired $\epsilon_S$ in the right hand panel.}
    \label{fig:stopping_precision}
\end{figure}

In \cref{fig:stopping_precision} we compare the precision in the kinetic energy of the projectile to the resultant precision in the stopping power. We find that for a precision of 0.1 eV/\AA~ in the stopping power we require a precision in the kinetic energy of approximately 0.02 Ha at the highest velocity which corresponds to approximately  $10^4$ samples. If the precision is lowered to 0.1 eV / \AA~ this sampling overhead drops to around $\mathcal{O}(10^2)$. 

\end{document}